\documentclass[10pt,preprint]{aastex}
\usepackage[utf8]{inputenc}
\usepackage[table]{xcolor}
\usepackage{rotating}

\newcommand\simlt{\lower.5ex\hbox{$\; \buildrel < \over \sim \;$}}
\newcommand\simgt{\lower.5ex\hbox{$\; \buildrel > \over \sim \;$}}

\begin{document}

\title{Can we reconcile the TA excess and hotspot with Auger observations?}
\author{Noemie Globus\altaffilmark{1}, Denis Allard\altaffilmark{2}, Etienne Parizot\altaffilmark{2}, Cyril Lachaud\altaffilmark{2}, Tsvi Piran\altaffilmark{1}}
\altaffiltext{1}{Racah Institute of Physics, The Hebrew University of Jerusalem, 91904 Jerusalem, Israel}
\altaffiltext{2}{Laboratoire Astroparticule et Cosmologie, Universit\'e Paris Diderot/CNRS, 10 rue A. Domon et L. Duquet, 75205 Paris Cedex 13, France}

\onecolumn
\small
The Telescope Array (TA) shows  a 20$^{\circ}$ hotspot as well as an excess of UHECRs above  50~EeV when compared with the Auger spectrum. We consider the possibility that both the TA excess and  hotspot are due to  a dominant source in the Northern sky. We carry out detailed simulations of UHECR propagation in both the intergalactic medium and the Galaxy, using different values for the intergalactic magnetic field. 
We  consider two general classes of sources: transients and steady, adopting  a mixed UHECR composition that is consistent with the one found by Auger. The spatial location of the sources is draw randomly.
We generate Auger-like and TA-like data sets from which we determine the spectrum, the  sky maps  and  the level of anisotropy.
We find that, while steady sources are favored over transients,  it is unlikely to account for all the currently available observational data. 
While we reproduce fairly well the Auger spectrum for the vast majority of the simulated data sets, most of the simulated data sets with a spectrum compatible with that of TA (at most a few percent depending on density model tested) show a much stronger anisotropy than the one observed. We find that the rare cases in which both the spectrum and the anisotropy are consistent require a steady source within $\sim 10$ Mpc, to account for the flux excess, and a strong extragalactic magnetic field $\sim 10$ nG, to reduce the excessive anisotropy. 
\twocolumn

\section{Introduction}

The origin of ultra-high-energy cosmic rays (UHECRs) is still unknown, despite intense theoretical studies and observational efforts. A new generation of detectors have collected, in the last decade,  data with unprecedented statistics in both hemispheres. 
The Pierre Auger Observatory \citep{AugerObs} (hereafter Auger), operating in Argentina since 2004, is the largest observatory, with $\sim3000$~km$^{2}$ effective detection area, and an integrated exposure of $66452\,\mathrm{km^{2}}\,\mathrm{sr}\,\mathrm{yr}$ as of March 31$^{\mathrm{st}}$, 2014, mostly in the Southern sky. Telescope Array \citep{TAObs} (hereafter TA), operating in Utah (USA) since 2007, covers $\sim700$~km$^{2}$, and obtained an integrated exposure of $8600\,\mathrm{km^{2}}\,\mathrm{sr}\,\mathrm{yr}$ as of May 11$^{\mathrm{th}}$, 2015, in the Northern sky.

The results reported by these two international collaborations concerning the energy spectrum, composition and angular distribution show some potentially significant differences. While a joint analysis working group concluded that the UHECR composition estimated by both experiments are compatible \citep{Unger15}, the TA and Auger spectra appear to be different in shape, and an intermediate-scale anisotropy has been reported by the TA collaboration, while the Auger skymap does not show a statistically significant deviation from isotropy at the highest energies, 
despite its larger exposure.
While the difference in the spectra might possibly be due to energy-dependent systematics, we explore here the possibility that one or several 
source(s) in the Northern sky is (are) responsible for both the excess in the high-energy spectrum and to the so-called hotspot in the TA data.

Although the conjunction of these two observations may appear natural 
at first sight, with a particularly intense and/or nearby source producing a larger flux than average as well as a distinct cluster of events, the quantitative study of this conjunction rather points towards a severe conflict: 
the excess in the TA flux compared to Auger is so large above $ 50$~EeV, that a much stronger anisotropy than the one observed is to be expected in general. 
While large deflections of the UHECRs by intervening magnetic fields may be invoked as a way to ``wash out'' the anisotropy, this seems essentially impossible if the flux is dominated by protons, and even in the case of heavier nuclei, large deflections tend to reduce the possibility of strong localised flux excesses, notably in the case of transient sources, because they spread the flux over larger time windows.

In order to quantify the problem at hand and to estimate the likeliness of such a combination of observations, we carry out detailed simulations of UHECR propagation in both the intergalactic medium (IGM) and the Galaxy, and distinguish between two general classes of sources:  transients and steady. {\color{black} We use a specific model for the sources, based on gamma-ray bursts (GRBs) \citep{Globus15a}, which reproduces both the spectrum and composition observed by Auger}. 

We first investigate,  in Sect.~\ref{sec:significance}, the significance of the difference between the Auger and TA measurements. We then present in Sect.~\ref{sec:analytical} a simple analytical estimate to determine the conditions under which a single source could contribute at a high level. In Sect.~\ref{sec:simulations} we discuss  the Monte Carlo simulations, where we take into account the propagation of the UHECRs from their sources to the Earth, including the model parameters and the procedure used to generate Auger-like and TA-like data sets from which sky maps can be built and analysed. We  present, in Sect.~\ref{sec:results}, the results of our simulations, applying definite criteria to assess the similitude or compatibility of the simulated data sets with the actual observational data. Finally, we summarise the results and discuss their possible implications in Sect.~\ref{sec:discussion}.

\section{Are the UHECR northern sky and southern sky significantly different?}
\label{sec:significance}

The possible difference in composition of the UHECRs observed by TA and by Auger has been widely discussed, after the claim by TA that their data is compatible with a pure proton composition (see for instance \cite{TACompo}, whereas Auger reports a gradual, but very significant trend towards higher mass nuclei around 10~EeV \cite{AugerCompo1,AugerCompo2}. 
After a joint analysis \citep{Unger15} the TA data is not inconsistent with the transition towards heavier elements inferred from the Auger data \citep[see however][ for a different interpretation of the data]{ShahamPiran13}.

The difference in the clustering of events as well as in the energy spectrum appears more striking, and apparently more significant. While no significant small or intermediate-scale anisotropy can be observed in the Auger data [the largest departure from isotropy was found to have a post-trial probability of $\sim1.4$\% \citep{Aab15}], the TA Collaboration reported a so-called hotspot, with a $20^{\circ}$ angular scale, near the constellation Ursa Major.
The chance probability of observing such a clustering anywhere in the sky is $3.7\cdot10^{-4}$, equivalent to a one-sided probability of $3.4~\sigma$ \citep{Tinyakov15}. 

While such a level of significance is too low to be conclusive, it should be considered together with another difference, regarding the energy spectrum above $\sim50$~EeV.
Fig.~\ref{fig:spectrumData} depicts the Auger and TA data, where a shift of $-13\%$ has been applied to the TA energy scale, as recommended by the Auger-TA joint working group \citep{Unger15}. The TA spectrum clearly shows a significant excess at higher energy, at least if one considers only the statistical error bars (shown on the plot). A systematic uncertainty with a rather strong energy dependence would be needed to explain such a difference.

After scaling down the energy by 13\% there are 83 highest energy TA events above 50~EeV.
They correspond to an exposure of 8600~km$^{2}\,\mathrm{sr}\,\mathrm{yr}$ \citep{matthews_talk}. On the other hand, Auger reports 231 events above 52~EeV, for an exposure of 66452~km$^{2}\,\mathrm{sr}\,\mathrm{yr}$. Given the shape of the spectrum between 50~and 60~EeV, this extrapolates to  $\sim 290$ events above 50~EeV. If the Auger flux is assumed to represent the average UHECR flux in the absence of anisotropy, then the expected number of events for TA is $\sim 38$. The actual integrated flux of TA would thus need to be a 7$\sigma$ upward fluctuation.

It thus appears unlikely that the UHECR fluxes observed by Auger and TA are just different realisations of an underlying roughly isotropic flux. 
Put together with the observation of the TA hotspot, the current data suggest the possibility that not only the apparent excess in the TA spectrum above 50~EeV is real (unless large energy-dependent systematic uncertainties impact the measurements), but also that this excess could be caused by the contribution of one (or more) localised sources, which would dominate mostly in the Northern sky. Quantitatively, if the integrated flux of Auger above 50~EeV represents an average contribution of typical sources distributed more or less isotropically over the sky, the corresponding contribution in the TA data should be $\sim 38\pm 6$~events, which leaves $\sim 45\pm 6$ for the putative additional source(s). Thus, if the difference between the two spectra is taken seriously and attributed to the contribution of a dominant source, this source may represent 45\%--60\% of the total Northern sky flux.

To this end, we do not limit ourselves to the consideration of the spectrum. A satisfying model must not only provide two different spectra in the Northern and Southern hemispheres, but also reproduce the anisotropy patterns: it must i) be compatible with isotropy in the Southern sky (i.e not produce an anisotropy signal much stronger than the warm spot reported around the direction of Cen A, with a 1.4\% post trial chance probability to arise as a fluctuation of an isotropic flux), and ii) provide a hotspot in the Northern sky with a typical angular scale of 20$^{\circ}$. 

Concerning the other observable of UHECR phenomenology, namely the composition, it is taken into account here in a generic way. In \citet{Globus15a}, some of us have developed a model based on the acceleration of particles in the mildly relativistic internal shocks of GRBs. This model  reproduces the spectrum and composition\footnote{in order to reproduce the observed UHECR flux we had to assume that cosmic-rays carry most of the energy dissipated at the internal shock especially in the case of low luminosity GRBs.}, both below and above the ankle \citep{Globus15b}. From a phenomenological point of view, the main features of this model are a low value of the maximum energy for protons at the sources, a hard source spectrum for all nuclei except protons (which have a significantly softer spectrum), and a source composition with a metallicity higher than the usual Galactic cosmic-ray component by a factor of $\sim 10$. These are considered here as generic features of a "working model''   , providing a suitable description of the average UHECRs, independently of the actual sources, whether GRBs, other types of transient sources (like tidal disruptive events, see e.g. \cite{Komossa15}), or steady sources. For the purpose of the anisotropy analyses of this paper, the main relevant ingredient is the composition of the UHECRs with an energy larger than 50~EeV, which is thus assumed to be the same as that of our explicit GRB model \citep{Globus15a}, but without prejudice regarding the nature of the sources.

\section{Analytical estimates}
\label{sec:analytical}

\subsection{Transient vs permanent sources}
\label{sec:tvsp}

We describe now a simplified analytical investigation of the conditions under which an individual source can be responsible for a significant fraction of the total observed UHECR flux 
above an energy $E_{\mathrm{th}}$ (e.g. $E_{\mathrm{th}} = 50$~EeV). 
We also estimate under the same conditions the angular size of this source in the sky.

As can be easily understood, the situation is different depending on the nature of the sources, whether transient or permanent. In both cases, the apparent angular size of a given source on the sky depends on the deflections of the particles by the intervening magnetic fields\footnote{we don't consider here the possibility that the source is extended and the observed size reflects its intrinsic scale.}. However, the apparent flux level of a source cannot be estimated in the same way for transient and steady sources. 
We can only see a transient source when it is active towards us, i.e. when the difference in time between the moment when we are observing it and the moment when it was actually active matches the particle propagation time. 
But the \textit{spread} in this propagation time 
 is what sets the instantaneous flux level. For a given total amount of energy emitted by the source,  
the apparent luminosity is simply inversely proportional to the spread of the time delays.

In practice, we consider as a transient source a source for which the typical variation of the intrinsic luminosity timescale is shorter than the propagation time delay from the source to the Earth. In this case, the effective luminosity is dictated by the total energy release during the transient event divided by the propagation time spread. We also assume, for the sake of simplicity, that the sources are standard candles. This is a conservative assumption, since a spread in the intrinsic source luminosities typically increases the variance of the contribution of the dominant source to the total flux, so the probability to obtain a higher flux fraction becomes larger \citep{Blaksley13}.

Three steps of propagation, associated with different types of deflections and time delays, can be distinguished: i) around the source, before the freshly accelerated UHECRs reach the IGM, 
ii) in the IGM between the source and our Galaxy, and iii) in our Galaxy, up to  Earth. For the analytical estimates below, we gather these contributions into one standardised step, corresponding to the propagation of particles through a homogeneous medium filled with an isotropic turbulent magnetic field with a coherence length $\lambda_{\mathrm{c}} \equiv \lambda_{\mathrm{Mpc}}$~Mpc and intensity $B_{\mathrm{rms}} \equiv B_{\mathrm{nG}}$~nG. This is a very crude assumption, but it gives an idea of the link between the deflections and the scale of the intervening magnetic fields. Note that the specific influence of the Galactic magnetic field (GMF) is analysed in more detail in the following sections, as it turns out to play an important role, especially in the case of transient sources.

In the limit of the weak scattering regime, where the Larmor radius of the cosmic-ray is larger than the turbulence scale of the magnetic field, the typical scale of the angular deflections of a particle with charge $Z$ and energy $E \equiv E_{20}\times 10^{20}$~eV, coming from a source at a distance $D_{\mathrm{s}} \equiv D_{\mathrm{Mpc}}$~Mpc, is given by \citep{WM96}:
\begin{equation}
\Delta\theta  \approx 0.36^{\circ} \, Z\,E_{20}^{-1}B_{nG} \, D_{\mathrm{Mpc}}^{1/2}\, \lambda_{\mathrm{Mpc}}^{1/2}.
\label{eq:deltaTheta}
\end{equation}
If the dominant class of nuclei at 50~EeV is CNO \citep{Globus15b}, 
the angular deflection is of the order of 14--18$^{\circ}$ in a nanogauss extragalactic magnetic field (EGMF) with a Mpc coherence length for a source located at 10~Mpc, or 7.5--10$^{\circ}$ for a source located at 3~Mpc. 
In comparison, the typical deflection in a 5~$\mu G$ GMF with a coherence length of 200~pc over 1~kpc, is of the same order, around 8--10$^{\circ}$.

We now turn to the conditions under which a source may contribute a given fraction of the total UHECR flux received on Earth. In principle, this fraction depends on energy, as the apparent source spectrum is expected to be different from the overall UHECR spectrum resulting from the contribution of all sources and all distances, 
shaped notably by the GZK horizon effect. We thus consider only the integrated flux above a given energy, $E_{\mathrm{th}}$ (e.g. $E_{\mathrm{th}} = 50$~EeV), and denote by $\eta_{\mathrm{flux}}$ the fraction of the total flux above that energy, which is contributed by the individual source. In all cases, we assume that the source emits UHECRs isotropically.

Note that an excess in the spectrum in some part of the sky could be due to the contribution of two sources located by chance in the same hemisphere, rather than to the contribution of just a single source. However, a simple reasoning shows that this is in general much less probable. 
Essentially, the flux from one source is typically reduced by a factor of 2 if its distance is increased by a factor of $\sqrt{2}$. Now, it appears that the chance probability for one source to contribute a substantial fraction of the UHECR flux is relatively low, of the order of a few percent in the favorable scenarios. This is related to the probability to find a source closer than some distance $D$, which we may write $\mathcal{P}_{1}(D)$. The probability to find two sources in the same hemisphere closer than the distance $\sqrt{2}\times D$, i.e. in a volume $2\sqrt{2}$ larger, is then given by $\mathcal{P}_{2}(\sqrt{2}D) \simeq \frac{1}{2}[\mathcal{P}_{1}(\sqrt{2}D)]^{2} \simeq 4\times \mathcal{P}_{1}(D)^{2}$. Since $\mathcal{P}_{1}(D)$ is much smaller than 1/4, this has a much smaller probability. Therefore, in the analytical exploration below, we consider only the case of one dominant source being responsible for the excess in the overall flux.

\subsection{Transient sources}
\label{sec:burstingSources}

In the case of bursting sources, the apparent flux of a source, $\Phi_{\mathrm{app}}$, is governed by its total energy (emitted in the form of UHECRs above $E_{\mathrm{th}}$), $E_{\mathrm{s}}$, its distance, $D_{\mathrm{s}}$, and the average time spread, $\Delta t$, over which it is visible from Earth in the considered energy 
range. It is given by:
\begin{equation}
\Phi_{\mathrm{app}} = \frac{E_{\mathrm{s}}}{4\pi D_{\mathrm{s}}^{2} \Delta t}.
\label{eq:PhiApp}
\end{equation}

This should be compared with the average UHECR flux on Earth above $E_{\mathrm{th}}$, $\Phi_{\mathrm{UHECR}}$, which can be approximately expressed in terms of the average source energy, $E_{0}$, the source rate, $R = R_{-9} \times 10^{-9}\,\mathrm{Mpc}^{-3}\mathrm{yr}^{-1}$, and the GZK horizon scale, $H = H_{100}\times 100$~Mpc, as
\begin{equation}
\Phi_{\mathrm{UHECR}} \approx E_{0}\,R\,H.
\label{eq:PhiUHECR}
\end{equation}

In the above-mentioned magnetic field configuration, the spread in the time delays is approximately given by \citep{WM96}:
\begin{equation}
\Delta t  \approx  (30 ~{\rm yr}) \, Z^{2}\, E_{20}^{-2} \,B_{nG}^{2} \, D_{\mathrm{Mpc}}^{2} \,\lambda_{\mathrm{Mpc}}.
\label{eq:timeSpread}
\end{equation}

Here we abusively use the expression for the average time delay of the particles, instead of its spread over an ensemble of particles. This is not strictly correct, and indeed in the detailed simulations discussed in the next sections, the flux level of the sources is determined by the actual propagation time of each individual particle, obtained through a full Monte-Carlo procedure, taking also into account the rigidity losses along the way. However, for the current analytical estimate, we keep the above formula which has the advantage of providing an explicit dependence in the various parameters. By this, we acknowledge the fact that, according to the simulations, the spread in the time delays of the bulk of the particles is indeed of the same order as the average time delay.

From Eqs.~(\ref{eq:PhiApp}) and~(\ref{eq:PhiUHECR}), we can simply write the ratio between the flux of the source under study and the total flux expected on average (over space and time in the local universe, i.e. over the so-called \textit{cosmic variance}
) above $E_{\mathrm{th}}$. This will be referred to below as the source-to-average-total (or STAT) flux ratio. It reads:
\begin{eqnarray}
\eta_{\mathrm{flux}} &\equiv& \frac{\Phi_{\mathrm{app}}}{\Phi_{\mathrm{UHECR}}}= \frac{E_{\mathrm{s}}}{E_{0}}\,\frac{1}{R\Delta t}\,\frac{1}{4\pi D_{\mathrm{s}}^{2}H} \nonumber\\&\approx& 2.7\,10^{4}  \xi \, R_{-9}^{-1} \, E_{20}^{2}\, Z^{-2}\, B_{\mathrm{nG}}^{-2} \, D_{\mathrm{Mpc}}^{-4} \,\lambda_{\mathrm{Mpc}}^{-1}\, H_{100}^{-1}\,,
\label{eq:etaFlux}
\end{eqnarray}
where we noted $\xi = E_{\mathrm{s}}/E_{0}$ the over-luminosity of the source under consideration. In the case of standard candles, $\xi = 1$.

Obviously, a very nearby source will give a very high flux fraction. However, $\eta_{\mathrm{flux}}$ depends very strongly on the source distance, as its fourth power, since $D_{\mathrm{s}}$ comes in both through the trivial $1/D_{\mathrm{s}}^{2}$ flux factor, and through the apparent time spread of the source, proportional to $D_{\mathrm{s}}^{2}$. Therefore, one needs to estimate the probability to find a source within a given source distance.

The average number of transient sources, active at any given time, in a sphere of radius $D_{\mathrm{s}}$, is $\left<N\right> = R \times \Delta t \times \frac{4}{3}\pi D_{\mathrm{s}}^{3}$. For low values of $\left<N\right>$, this is also the probability to find a source within $D_{\mathrm{s}}$.
Note that we are implicitly interested in situations where this number is low indeed, since we are interested in the dominant source, whose flux compares to the total flux of all other sources contributing in that energy range. It must thus be exceptional by its proximity and/or small time spread, which would not be the case if $\left<N\right>$ were close to 1 or more.

Equation~(\ref{eq:etaFlux}) can now be inverted to give the distance at which a source with over luminosity $\xi$ must be located to provide a given STAT ratio, $\eta_{\mathrm{flux}}$:
\begin{eqnarray}
D_{\mathrm{s}}(\eta_{\mathrm{flux}}) \simeq (12.8\,\mathrm{Mpc}) \, \xi^{1/4} \,  E_{20}^{1/2} \, (Z B_{\mathrm{nG}})^{-1/2} \nonumber\\ \cdot\,  ( \eta_{\mathrm{flux}}\,R_{-9} \lambda_{\mathrm{Mpc}} \, H_{100})^{-1/4}.
\label{eq:distance}
\end{eqnarray}

The corresponding probability, for $\xi = 1$, is then found to be:
\begin{eqnarray}
\mathcal{P}(\eta_{\mathrm{flux}}) &\simeq& \left<N(D_{\mathrm{s}}(\eta_{\mathrm{flux}})\right> \nonumber\\&\simeq& 4.3\% \,\, \eta_{\mathrm{flux}}^{-5/4} \, E_{20}^{1/2} (Z B_{\mathrm{nG}})^{-1/2}  \, R_{-9}^{-1/4} \, \lambda_{\mathrm{Mpc}}^{-1/4}\, H_{100}^{-5/4}.\nonumber\\ 
\label{eq:probaEtaFluxBurst}
\end{eqnarray}
Let us consider CNO nuclei at $E \simeq 50$~EeV and a source rate of one burst per Gpc$^{3}$ per year. 
If the deflexions and time delays are similar to what would produce a homogeneous turbulent EGMF with $B_{\mathrm{nG}} = \lambda_{\mathrm{Mpc}} = 1$, the probability that a dominant source contributes at the same level as the rest of all other sources in one hemisphere, i.e. $\eta_{\mathrm{flux}} \simeq 1/2$ is of the order of 3\%. Although not common, this does not seem particularly unlikely. However, one must also consider the angular extension of such a source.

Using Eq.~(\ref{eq:deltaTheta}), one obtains the angular size expected for a source that would contribute a fraction $\eta_{\mathrm{flux}}$ of the global UHECR flux, i.e. be located at distance $D(\eta_{\mathrm{flux}})$, given by Eq.~(\ref{eq:distance}):
\begin{eqnarray}
&&\Delta\theta(\eta_{\mathrm{flux}}) \nonumber\\&&\simeq 1.3^{\circ} \, \eta_{\mathrm{flux}}^{-1/8} \, \xi^{1/8} \, E_{20}^{-3/4}\, (ZB_{\mathrm{nG}})^{3/4} \lambda_{\mathrm{Mpc}}^{3/8} \, (R_{-9}H_{100})^{-1/8}.\nonumber\\ 
\label{eq:angularSize}
\end{eqnarray}

As can be seen, the dependence on the various parameters is weak, except on the ratio $E/ZB$, which is simply the Larmor radius of the particles (times $c/e$). For CNO nuclei at 60~EeV in a nanogauss field, one obtains $\Delta\theta \simeq 7$--$9^{\circ}$. This appears too small to account for the TA observations. As a matter of fact, a source that would contribute, say, 50\%, of the total flux and would span over 9$^{\circ}$ in the sky.
Therefore, if the apparent difference in the UHECR spectrum between the northern and southern hemispheres is to be attributed to the contribution of one dominant source, then the angular spreading of the source must be due mostly to the action of the GMF, at least if the extragalactic magnetic field (EGMF) does not significantly exceeds a nanogauss. For a pervading EGMF of 3~nG, say, the angular size would increase by a factor 2.3, while the probability of a given contribution to the overall flux would decrease by a factor 1.7 (see Eq.~(\ref{eq:probaEtaFluxBurst})). Note that the angular size of a source could be increased if the source is surrounded by a large magnetic halo, which by itself has a large angular extension over the sky. However, in the case of transient sources, such a halo would also considerably extend the time spread of the UHECRs from that source, and thus strongly reduce its apparent luminosity.

A key question is thus whether the GMF can increase the angular spreading of a source sufficiently to reconcile the idea that the spectral differences between the two hemispheres are due to a dominant source with the anisotropy data. Unfortunately, the GMF is not known with enough precision to give a definite answer to this question. We discuss this point below in more detail, in particular with the comparison between the situations in the northern and southern hemispheres, which turn out to be different, but we recall here the estimate of Sect.~\ref{sec:tvsp}, which gave $\sim 8-10^{\circ}$ for a length of 1~kpc through a 5~$\mu$G with $\lambda_{\mathrm{c}} = 200$~pc. Obviously, the deflections depend on the length of the trajectory through the GMF, and thus on the direction of the source, within or away from the Galactic disk. In addition, the size of the magnetic halo above the disk appears to be a crucial parameter (as well as its coherence length). An extension much larger than 1~kpc, still with significant magnetic field, would lead to significantly larger deflections, and could explain the absence of a well marked, small scale anisotropy, even if a large fraction of the total flux can be attributed to one source with relatively limited angular spread at the entrance of the Galaxy.

Finally, it is instructive to calculate the typical time spread of the UHECRs detected from a dominant source with STAT ratio $\eta_{\mathrm{flux}}$. This is easily obtained as a function of the various parameters, by reporting the required source distance, $D_{\mathrm{s}}(\eta_{\mathrm{flux}})$, given by Eq.~(\ref{eq:distance}), into the expression of $\Delta t$ in Eq.~(\ref{eq:timeSpread}). One finds:
\begin{eqnarray}
\Delta t (\eta_{\mathrm{flux}}) \simeq (4.9\,\mathrm{kyr})\times \eta_{\mathrm{flux}}^{-1/2} \xi^{1/2} Z E_{20}^{-1} B_{\mathrm{nG}}\lambda_{\mathrm{Mpc}}^{1/2} H_{100}^{-1/2}.\nonumber\\ 
\label{eq:deltaT}
\end{eqnarray}

For CNO at 60~EeV, the typical time spread is thus of the order of a few tens of kiloyears for the dominant source. It can be seen that a larger EGMF allows for a larger time spread, but only linearly, whereas $\Delta t$ increases as the square of the magnetic field for a given source distance. Thus, the larger value of $\Delta t$ allowed for a larger magnetic field is actually related to the necessity of having a closer source, which then reduces the probability of such an occurence (see Eq.~\ref{eq:probaEtaFluxBurst}).

\subsection{Steady sources}

We now turn to the case of steady sources, which inject UHECRs in the intergalactic medium at a constant luminosity, $L_{0}$. Like before, we allow for a different luminosity of the source under consideration, $L_{\mathrm{s}}$, and note $\xi \equiv L_{\mathrm{s}}/L_{0}$. The flux received from that source is simply:
\begin{equation}
\Phi_{\mathrm{app}} = \frac{L_{\mathrm{s}}}{4\pi D_{\mathrm{s}}^{2}},
\label{eq:PhiApp2}
\end{equation}
while the average UHECR flux from all sources within the GZK horizon, $H$, is:
\begin{equation}
\Phi_{\mathrm{UHECR}} \approx L_{0}\,n_{\mathrm{s}}\,H,
\label{eq:PhiUHECR}
\end{equation}
where $n_{\mathrm{s}} \equiv n_{-5} \, 10^{-5}\,\mathrm{Mpc}^{-3}$ is the source density.

The STAT flux ratio is thus given by:
\begin{equation}
\eta_{\mathrm{flux}} = \frac{\xi}{4\pi D_{\mathrm{s}}^{2}Hn_{\mathrm{s}}} \simeq 80 \,\, \xi \, D_{\mathrm{Mpc}}^{-2}\,H_{100}^{-1}\,n_{-5}^{-1}.
\label{eq:etaFluxSteady}
\end{equation}

The distance where a source must be to account for a flux ratio $\eta_{\mathrm{flux}}$ is thus:
\begin{equation}
D(\eta_{\mathrm{flux}}) = (8.9\,\mathrm{Mpc}) \,\, \eta_{\mathrm{flux}}^{-1/2} \, \xi^{1/2} \, n_{-5}^{-1/2} \, H_{100}^{-1/2}.
\label{eq:dist}
\end{equation}

As in the case of transient sources, the probability to find a source within this distance is essentially the average number of sources in the corresponding volume, namely $\left<N\right> = \frac{4}{3}\pi D_{\mathrm{s}}^{3} \times n_{\mathrm{s}}$. This gives:
\begin{equation}
\mathcal{P}(\eta_{\mathrm{flux}}) \simeq 3.0\% \,\, \xi^{3/2} \, \eta_{\mathrm{flux}}^{-3/2} \, n_{-5}^{-1/2} \, H_{100}^{-3/2}.
\label{eq:probaEtaFluxSteady}
\end{equation}

Regarding the angular spread of such a source, it is estimated as before using Eq.~(\ref{eq:deltaTheta}):
\begin{equation}
\Delta\theta(\eta_{\mathrm{flux}}) \simeq 1.1^{\circ} \,\, \eta_{\mathrm{flux}}^{-1/4} \, \xi^{1/4} \, Z E_{20}^{-1} \, B_{\mathrm{nG}} \,\,\lambda_{\mathrm{Mpc}}^{1/2} \,\, n_{-5}^{-1/4} \, H_{100}^{-3/4}.
\label{eq:angularSizeSteady}
\end{equation}

The dependence on the various parameters is larger than in the case of transient sources, because the flux level does not depend on any time spread, and thus on the magnetic field. However, the source angular extension does not depend strongly on the source density, which has a greater influence on the probability of the situation under investigation. For CNO nuclei at 60~EeV in a nanogauss magnetic field, we obtain an angular size of the order of 10--15$^{\circ}$, which is again too low to account for the TA hotspot. However, a stronger magnetic field can increase this angular size linearly, and a contribution of the GMF may also be important. Note that in a case of a pure proton composition scenario, the angular size of a hotspot would be smaller than $\sim3^{\circ}$ for a nanogauss magnetic field and a source distance of 10 Mpc. Thus it would be even more difficult to account for the anisotropy observed by TA.

\section{Model and simulations}
\label{sec:simulations}

In the previous section, we estimated the conditions under which one source may contribute a significant fraction of the total UHECR flux, and the probability that this situation occurs at a given moment in time, in a given location of the universe. The main outcome is that this probability may be of the order of a few percent for a relatively wide range of parameters, in the case of steady sources as well as in the case of transient sources. Concerning the angular size of this source on the sky, it appears that the EGMF alone is usually too low to spread the particles over a region much larger than 10--15$^{\circ}$ (CNO nuclei at 60~EeV), so that such a source would produce significant small-scale anisotropies, at variance with the current observations. However, larger values of the EGMF in the direction of the source or a significant contribution of the GMF to the overall deflections could in principle allow the scenario under study to be viable.

In this section and the next, we use a complete Monte Carlo simulation of the source distribution and history and of the UHECR propagation to study the transient and steady source scenarios in more detail, taking into account the GMF and the actual exposure of Auger and TA in their respective part of the sky, and comparing our simulated sky maps and spectra with the data. {\color{black}The overall compatibility of these sky maps with the actual data is discussed in Sect.~\ref{sec:results}}. Here, we describe the central numerical tool of the simulation, base on our Monte-Carlo propagation code, which is used to compute the propagation of UHECRs from individual point-like sources, taking into account their energy losses, their nuclear transmutation through the interaction with the cosmological photon background, and their deflections in the intervening magnetic fields
\citep[see][for more details about our propagation code]{Allard05,Globus08}.

\subsection{General ingredients of the simulations}
\begin{figure*}[t!]
\begin{center}
\includegraphics[width=0.33\linewidth]{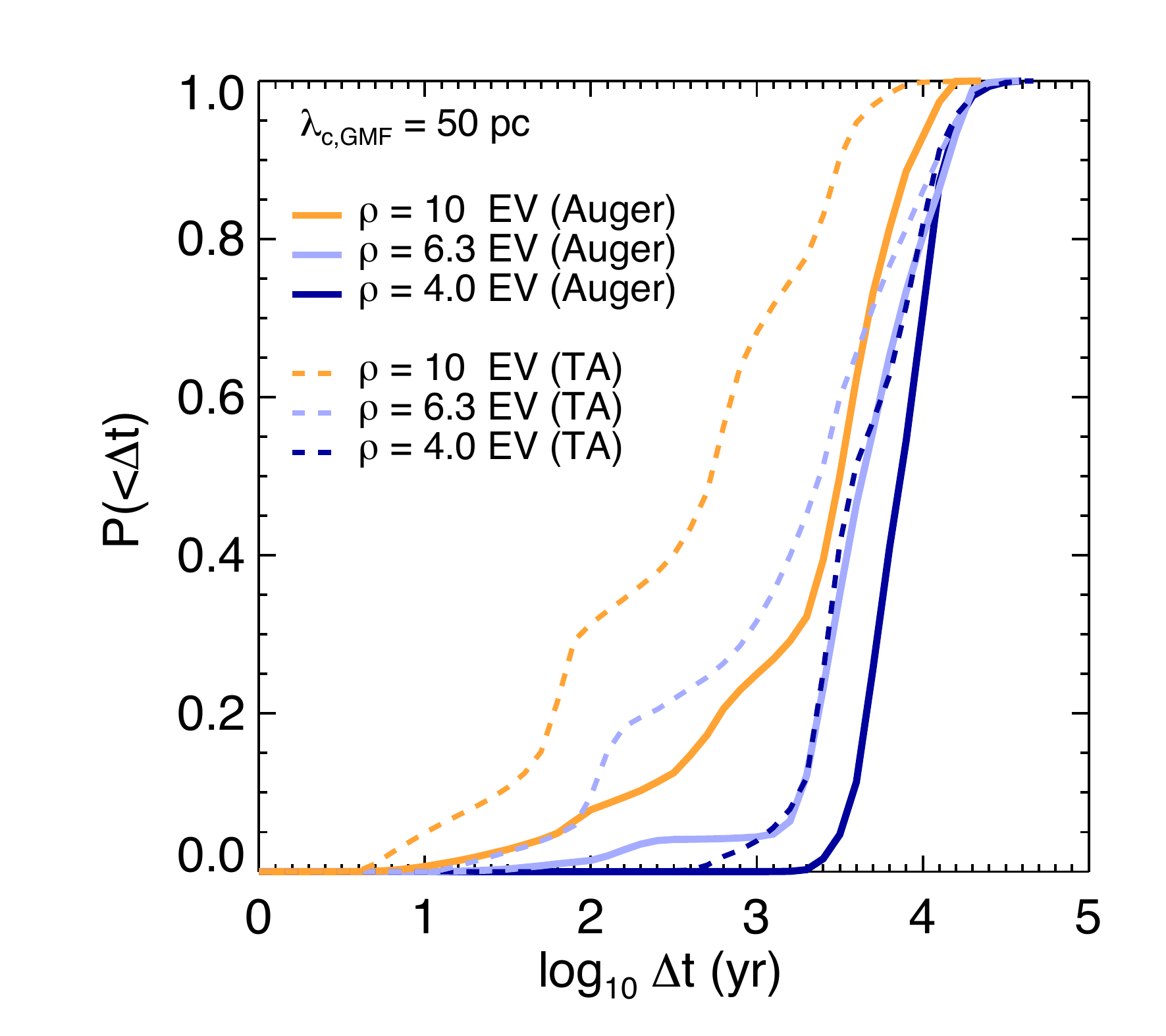}\includegraphics[width=0.33\linewidth]{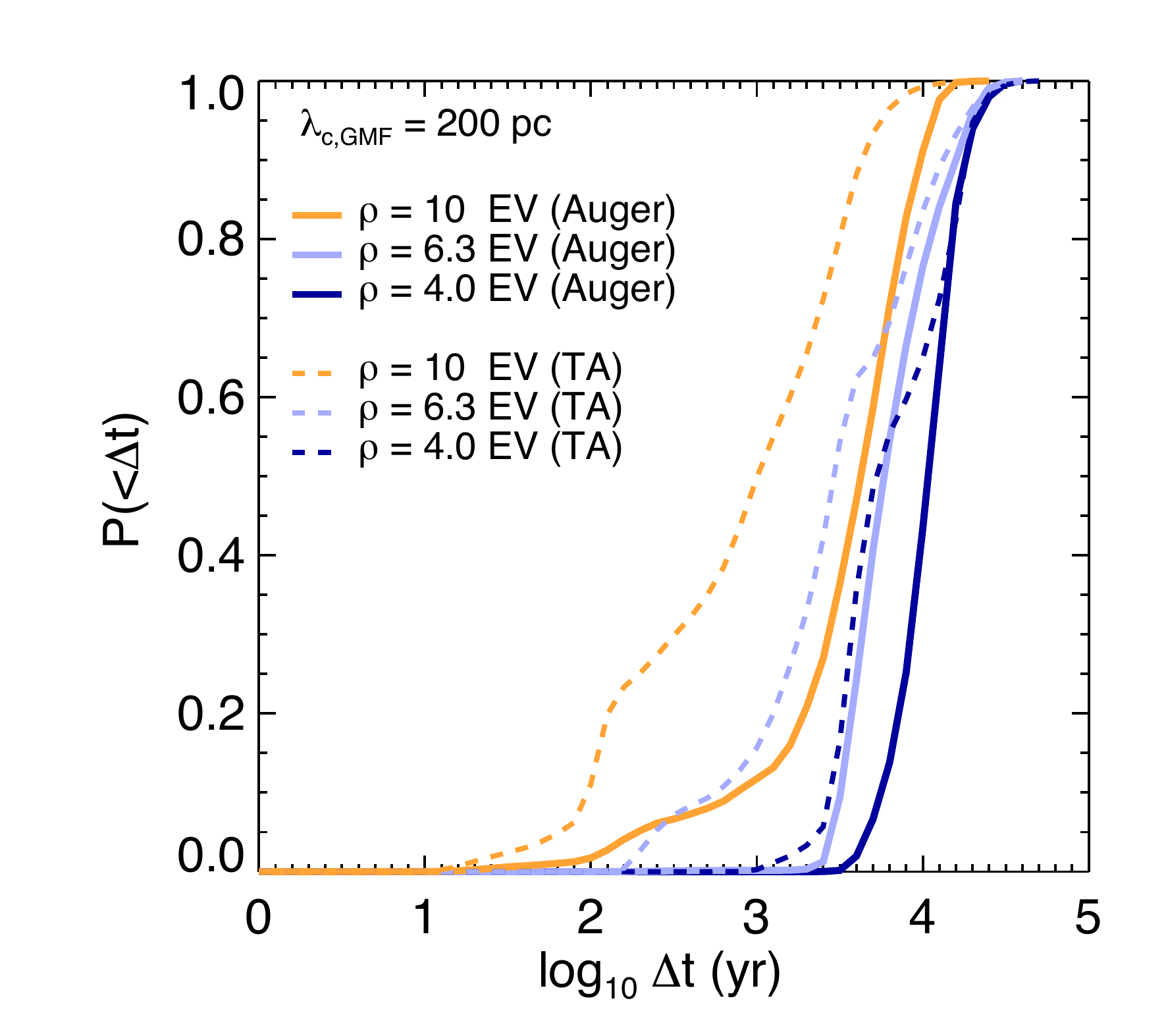}\includegraphics[width=0.33\linewidth]{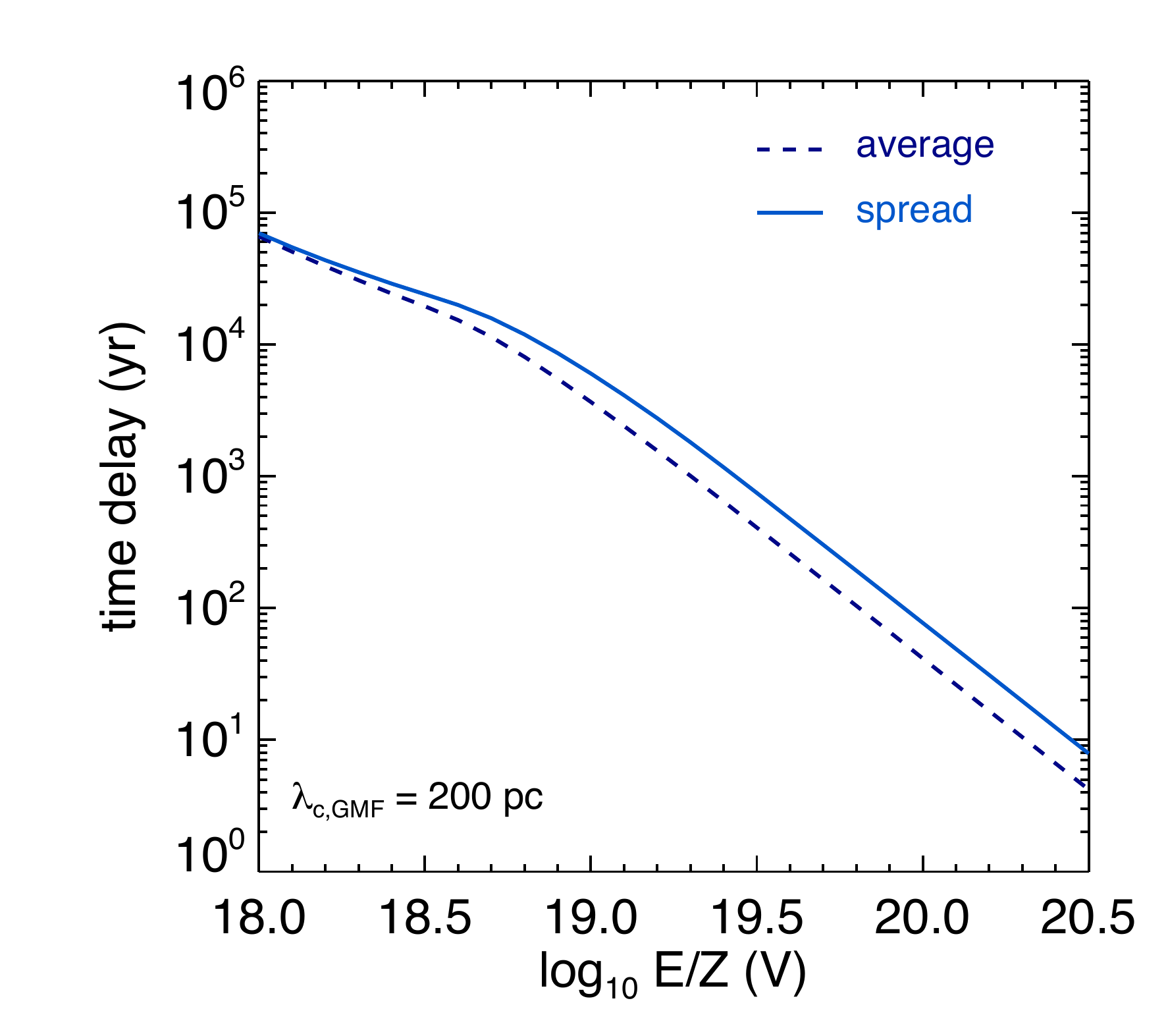}
\caption{Left and middle: cumulative distributions of the spread in time delays due to the GMF in the TA sky (dashed lines) and in the Auger sky (solid lines) for  different values of the rigidity, for $\lambda_{\mathrm{c,GMF}}=50$~pc (left) and for $\lambda_{\mathrm{c,GMF}}=200$~pc (middle). The ordinate shows the probability that the time delay is shorter than the value in abscissa. Note that the TA time delays are significantly shorter.
Right panel: average over all (incoming or outgoing) directions of the time delays and their spread due to the GMF, as a function of the rigidity of the particles. At high rigidities, they both scale as  $\Delta t\propto(E/Z)^{-2}$, as expected in the weak scattering regime (see Eq.~\ref{eq:timeSpread}).}
\label{fig:timeDelaySpread}
\end{center}
\end{figure*}

\subsubsection{Galactic and extragalactic magnetic fields}
\label{sec:GMFEGMF}

 We first follow the trajectory of individual particles through the EGMF, assuming a random Kolmogorov-like turbulent field with r.m.s. value $B_{\mathrm{EGMF}}$, and a principal turbulent scale of 1~Mpc (which corresponds to a coherence length $\lambda_{\mathrm{c}} \simeq 0.2$~Mpc). To model the EGMF, we follow the numerical procedure of \cite{GiacaloneJokipii99}.  The EGMF is assumed to be distributed homogeneously throughout the universe. Although this is admittedly not realistic, it allows us to explore the general effect of an EGMF.  
As can be noted from the analytical study of Sect.~\ref{sec:analytical}, the magnetic field value and coherence length always appear through the product $B_{\mathrm{EGMF}} \, \lambda_{\mathrm{c}}^{1/2} = (1\,\mathrm{nG}\,\mathrm{Mpc}^{1/2} )\,\mathrm{B}_{\mathrm{nG}}\,\lambda_{\mathrm{Mpc}}^{1/2} $. Therefore 
it is not necessary to vary these two parameters separately. We chose to keep the coherence length fixed, and simply vary $B_{\mathrm{EGMF}}$, in the range from 0.01~nG to 10~nG. For instance, the case corresponding to $B_{\mathrm{EGMF}} = 10$~nG and the default value of $\lambda_{\mathrm{c}} = 0.2$~Mpc is equivalent to a case with $B_{\mathrm{EGMF}} \sim4.5$~nG and $\lambda_{\mathrm{c}} = 1$~Mpc.

{\color{black}Concerning the propagation in the Galaxy, the large-scale structure as well as the so-called striated component of the GMF are modeled as in \cite{JF12} (hereafter JF12).  We add a purely turbulent component following the numerical procedure of \cite{GiacaloneJokipii99}, assuming a Kolmogorov-like spectrum and a coherence length $\lambda_{\rm c,GMF}$ of either 50~pc or 200~pc \citep{2016JCAP...05..056B}. The r.m.s. value follows the magnitude of the regular component with an overall enhancement factor of 3. The resulting local value of the GMF (near the Earth) is consistent with the 6 $\mu$G value inferred from observations \citep{Beck08}.}

We also consider the possible presence of a magnetic field around the source itself. In the absence of a well defined model, we simply assume here that the sources are embedded in a magnetic field similar to that of our own Galaxy.

\begin{figure}[t!]
\begin{center}
\includegraphics[width=\linewidth]{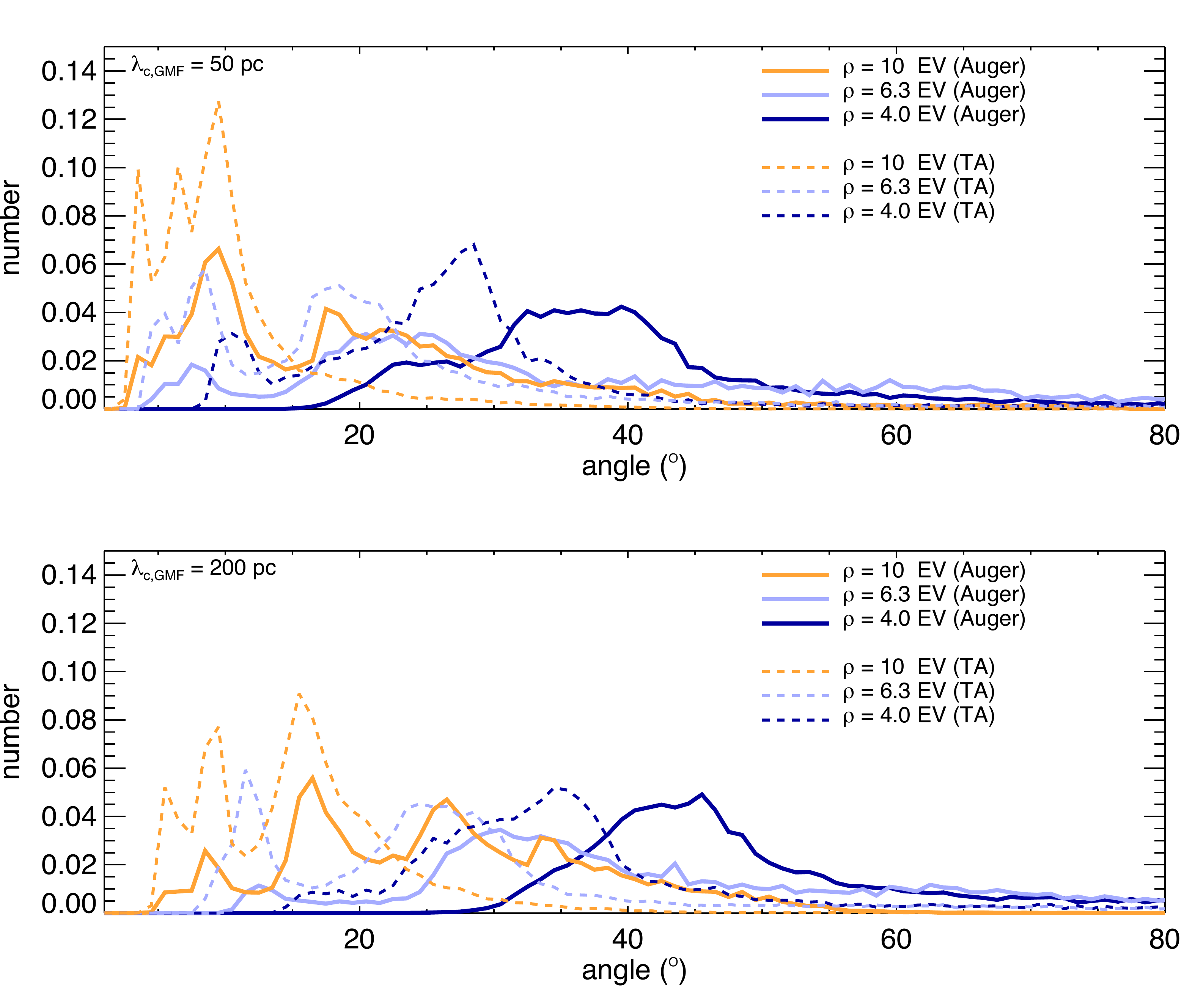}
\caption{Normalised histograms of the angular spreads of the sources visible in the TA sky (dashed lines) and in the Auger sky (solid lines), for different values of the rigidity. 
Top panel:   $\lambda_{\mathrm{c,GMF}}=50$ pc; bottom panel: $\lambda_{\mathrm{c,GMF}}=200$ pc.
The ordinate shows the probability that a given source be located in a direction of the sky for which the variance of the angular shift between the position of the source and the actual arrival direction of an observed UHECR from that source is equal to the value in abscissa.
}
\label{fig:angularSpread}
\end{center}
\end{figure}

\subsubsection{Effect of the Galactic magnetic field}
\label{GMFEffect}

The GMF becomes important  when the EGMF is typically smaller than $\sim0.3$~nG. In these cases its contribution is significant and may even dominate the deflection and the corresponding time delays, as far as the closest sources are concerned. We recall that the typical deflection of particles at rigidities 8-9 EV is smaller than $10^{\circ}$ in the extragalactic nanogauss field for a source located at 10 Mpc (see Eq. \ref{eq:angularSize}). At these rigidities the deflections by the GMF are of the order of $10^{\circ}$ to $20^{\circ}$, depending on the arrival direction of the particle.

The GMF has several important effects. First, it deflects the particles and makes a given source appear larger on the sky. This effect adds to the deflections through the EGMF. Second, it may cause some magnification or demagnification of the sources, depending on the arrival direction of the particles in the Galaxy (see for instance \cite{Harari99}). Indeed, a small solid angle around some specific directions in the sky (as observed ``backwards'' from the Earth) may gather the particles coming from a larger (respectively smaller) solid angle at the entrance of the Galaxy, so that the sources located in these directions can appear with a larger (resp. smaller) flux than if no deflections applied. Like all effects related to the magnetic field, this effect  depends on the rigidity of the particles, i.e. it may select a given mass range at a given energy, or a given energy range for a given type of nuclei. This effect was discussed in \cite{Benji14}, where magnification maps were also shown, so we do not discuss it further here. However, we recall that our simulations take it into account in a consistent way, by applying the relevant magnification factor to the flux of each individual source in the sky, depending on the arrival direction of their UHECRs.

The third effect of the GMF is that it introduces some time delay, in addition to that caused by the EGMF. This effect is very important in the case of transient sources, since the spread in the time delays has a direct incidence on the apparent flux of a given source (see Sect.~\ref{sec:analytical}). Now, it is very interesting to note that the Northern and Southern hemispheres are not equivalent in this respect, due to the asymmetry of the GMF, notably the different northern and southern extensions of the Galactic halo. Smaller dispersions in the time delays are significantly more common in the part of the sky observed by TA than in the part of the sky observed by Auger. This result was obtained from a systematic study, where we divided the sky in 49152 pixels covering identical solid angles, and determined the variance of the time delays caused by the GMF for a large number of particles (256 on average, depending on the magnification) observed in the corresponding directions. 

In Fig.~\ref{fig:timeDelaySpread}, we show the cumulative distribution of the spread in time delays, 
assuming a random source position with equal probability in any direction of the sky. This probability is shown by the solid line curves in the case of UHECRs observed by Auger, while it is shown by the dashed line curves in the case of TA. We show the results for three different values of the rigidity, namely $10^{19}$~V, $10^{18.8}$~V and $10^{18.6}$~V. The first one corresponds for instance to C nuclei at 60~EeV (or O nuclei at 80~EeV or protons at $10^{19}$~eV), while the last one corresponds to Fe nuclei at 100~EeV. 
Fig.~\ref{fig:timeDelaySpread} clearly shows that the probability of a short time delay, 
is much larger if the source is mostly seen in the TA sky.  
This difference in the ``northern'' and ``southern'' time spreads introduced by the GMF is only relevant when the values are larger than the time spreads introduced by the EGMF. 
We find that this is essentially the case for EGMF values smaller than 0.3~nG, which is consistent with the EGMF-induced time delay estimated for the dominant source in Eq.~(\ref{eq:deltaT}). The average value of the time delays in the GMF and their spread over all possible directions is shown in the right panel of Fig.~\ref{fig:timeDelaySpread} as a function of the rigidity of the particle. At high rigidities, they both scale as  $\Delta t\propto(E/Z)^{-2}$, as expected in the weak scattering regime (see Eq.~\ref{eq:timeSpread}).

The structure of the GMF introduces a similar difference regarding the angular spread of the sources, as shown in Fig.~\ref{fig:angularSpread}. The plots show the variance of the particle deflections in the same conditions, i.e. for particles observed in the TA sky (dashed lines) and in the Auger sky (solid lines), for three values of the rigidity and two values of the field coherence length. The most probable angular spreads are typically smaller in the TA sky than in the Auger sky. 

Finally, while a small spread in the UHECR time delays translates into a high instantaneous flux in the case of transient sources, it should be noted that additional time delays are likely to originate from the magnetic field in the environment of the source itself, which could be that of the host galaxy or extend from the scale of a local circumstellar environment up to the scale of a galaxy cluster (see the discussion in \cite{TakamiMurase2012}). In our simulations, we simply assume that the typical time delays associated with the exit of the UHECRs from their sources to the general intergalactic medium is similar to those introduced by our own galaxies. Since the orientation of the host galaxy with respect to the line of sight (and to a possible source axis, in the case of beamed ejection of UHECRs) is a priori random, we draw randomly in the distribution of time spreads among different directions across the host galaxy. 

\subsubsection{General source properties}

To build simulated sky maps corresponding to a given astrophysical scenario, we first need to choose the location of the sources. In the case of steady sources, we simply draw randomly the source locations in 3D space according to the source density, with equal probability in all directions of the sky. Although matter is not distributed uniformly in the local universe, this simplification has no significant incidence on our results, which are analysed from the point of view of their total energy spectrum and their intrinsic anisotropy, at intermediate angular scales, rather than their correlation with specific classes of sources or matter distribution. Note that the source density specified in each case is actually a comoving density (i.e. with a fixed average source number per comoving volume) and that we also allow for a possible evolution of the individual source power as a function of redshift. This has no significant impact on the spectra and anisotropy patterns in the GZK energy range, given the short horizon scale. In the case of transient sources, we also need to specify the time of their occurence, which is chosen randomly according to a certain rate.

For definiteness, we choose the GRB source model described in \citet{Globus15a}, where the UHECRs are accelerated at the internal shocks of GRBs, as a reference model. 
From this model, we borrow some characteristics, which may be considered as generic properties that typical source models must have to reproduce the Auger data, such as a mixed composition with high metallicity and a hard spectrum with low proton maximum energy. The same is assumed here for all sources, whether transient or steady. {\color{black}Note that similar characteristics have been found in \citet{Allard12, Taylor14, unger15a, Matteo15}. Other recent discussions of the potential astrophysical implications of UHECR data can be found, for instance,  in \cite{Aloisio14,Taylor15}.}


Many properties may in principle vary from one class of sources to the other, none of which can however be directly inferred from the current data. For instance, while the average injection rate density, expressed in $\mathrm{erg}^{-1}\mathrm{Mpc}^{-3}\mathrm{yr}^{-1}$, is well constrained by the observed flux, we do not know the total energy injected in the form of UHECRs by individual sources, nor even its average, which is directly related to the source density in the case of steady sources, or to the source occurence rate in the case of transient sources. Until the sources are identified, these can essentially be regarded as free parameters in the general modelling of the UHECRs. In addition, it should be kept in mind that all sources may not have the same luminosity. In the absence of better-motivated assumptions, we adopt the relative luminosity distribution that results from the calculations in the framework of our GRB source model, and apply it throughout this paper, including in the case of steady sources.  
This is thought to be more realistic than the standard candle assumption. The corresponding transient source occurrence rate is $1.3\,10^{-9}\,\mathrm{Mpc}^{-3}\mathrm{yr}^{-1}$ \citep{WP10}. We also explored 10 times larger and 10 times smaller rates, but found that it did not change the outcome of the study significantly (in conformity with the analytical estimates, where the rate appears with a power 1/8 as far as the apparent angular size is concerned, and 1/4 regarding the flux excess probability, see Eqs.~\ref{eq:probaEtaFluxBurst} and~\ref{eq:angularSize}).

 Note that we consider sources emitting UHECR isotropically. This includes the case of transient sources although there are observational evidence and energetic arguments favoring a beamed emission from GRBs  \citep{Frail01}. While the assumption of an isotropic assumption has essentially no impact on our results  for values of the EGMF of $\lesssim 1$ nG (at least as long as the contributions of the strongest sources are concerned), this no longer true for larger values of the EGMF. Moreover, in the case of a beamed emission of UHECRs, the flux received from a given GRB event could also be affected by the angular spreading of the jet in the immediate vicinity of the source depending on the magnetic environment of the host galaxy and on the exact value of the beaming angle. Ignoring these potential complications, the predictions we make especially regarding the probability of a strong flux excess in the TA sky have to be considered as relatively optimistic, all the more when the assumed EGMF is of 1 nG or larger. Even in the framework of our simplifying hypotheses we will see that it is extremely difficult to account for the present observational data with a transient source scenario.

\subsection{Simulation procedure and analysis}

\subsubsection{Production of the simulated data sets}
\label{sec:datasetProduction}

We build the sky maps by picking individual UHECR events, one after the other, until we reach the required number to match the integrated flux seen by Auger above a reference energy of 5 EeV, namely 59500 events (as of March 31, 2014, for a total exposure of 66452 km$^2$ sr yr). Although chosen arbitrarily, this energy is low enough for the UHECR flux to be considered essentially isotropic at this energy, and the corresponding number of events is large enough to not suffer from significant statistical fluctuation (including cosmic variance) in the Auger data. 

The way we draw the individual events reflects the complexity of the underlying phenomenology. In the case of transient source, we first calculate numerically the all-particle spectrum for a given realisation of the GRB history in the universe, taking into account galactic and extragalactic time delay distributions {\color{black} and all the relevant energy loss processes for protons and nuclei (see \citet{Globus08} and \citet{Benji14} for more details)}. This calculation allows us to build probability tables for the energy distribution, the redshift distribution at a given energy, the discrete sources contribution at a given redshift and a given energy and for the different sources contributing significantly the probability distribution of the different species and the associated distributions of deflexion angles. 
Our data sets are then produced by sampling successively these probability tables. We first draw the energy $E$ of the particle, then we draw the redshift at which the particle was injected by its source in the intergalactic medium. Given the redshift, we then pick up an actual source, choosing randomly among the possible ones, in case there are several in the corresponding redshift range for the specific realisation of the distribution of sources under study. We then draw randomly the mass and charge of the particle. From there, we pick the deflection angle in the EGMF for a particle with the identified energy and charge coming from this particular source. Given the position of the source, we then know the arrival direction of the particle in the Galaxy.

At this stage, the study of Galactic deflections comes in. By inverting the result of a massive backwards propagation calculation [see the description of the method in \cite{Benji14}], we obtained for each arrival direction outside the Galaxy 
the probability of being seen on Earth from a given direction in the sky. We can thus draw an actual arrival direction accordingly.

The final steps consist in two acceptance/rejection draws, to take into account the magnification factor (we draw a random number between zero and the maximum magnification factor at the relevant rigidity over the whole sky) and the exposure of the experiment under consideration (Auger, TA, or a hypothetical full-sky,  uniform exposure experiment). 
If the event is kept in these last steps, we attribute a reconstructed energy by applying a random Gaussian relative error to each event with a width of 15\% in the case of Auger, and of 20\% in the case of TA.  The event is then included in the dataset and the procedure is repeated until the number of events with reconstructed energies above 5~EeV is 59500 in the case of Auger. This gives in average $\sim$ 7700 events above 5~EeV in the case of TA as expected from the ratios of the time integrated exposures of the two observatories. The full dataset is used to calculate the observed spectrum while the 231 (respectively 83) highest energy events are selected in the Auger (respectively TA) sky to compare their intrinsic anisotropy to that of the observational data. 

In the case of steady sources, the procedure is exactly the same, except for the random draw of the redshift of the source,
since the individual sources are assumed to be permanent and their contribution to the present flux is integrated over redshifts (i.e. look-back times).

\subsubsection{Exploration of the cosmic variance, statistical fluctuations and systematic differences}
\label{sec:cosmicVariance}

Given the assumptions summarised above, the free parameters of the models are the EGMF value, the GMF coherence length and the source density or occurence rate. For each set of these parameters, the observed spectra and the distribution of the UHECR events over the sky may be very different, depending on the actual position of the sources, and their occurrence time in the case of transient sources. The range of these variations is usually referred to as the cosmic variance. In order to explore these various possibilities, the procedure is repeated several times for each choice of the model parameters.  Specifically, for each model we produce 1200 random realisations of the GRB explosion history, and 600 random realisations of the the source distribution in the universe in the case of steady sources. Then, for each of these realisations, we produce 10 random data sets with the intended statistics (i.e. 59500 events above 5~EeV in the Auger sky), whose differences reflect statistical fluctuations of the same underlying sky map, energy spectrum and composition. Thus, in total, we obtain 12000 and 6000 realisations of each individual astrophysical scenario.
For this, we use general criteria based on the corresponding energy spectrum and the analysis of the intrinsic anisotropy of the data sets. The latter analysis is performed on sky maps which are built, for each realisation, with the same number of events as the data to which it is compared (i.e. with the 83 highest energy events in the case of the TA, and with the 231 highest energy events in the case of Auger).

\begin{figure}[t]
\begin{center}
\includegraphics[width=\columnwidth]{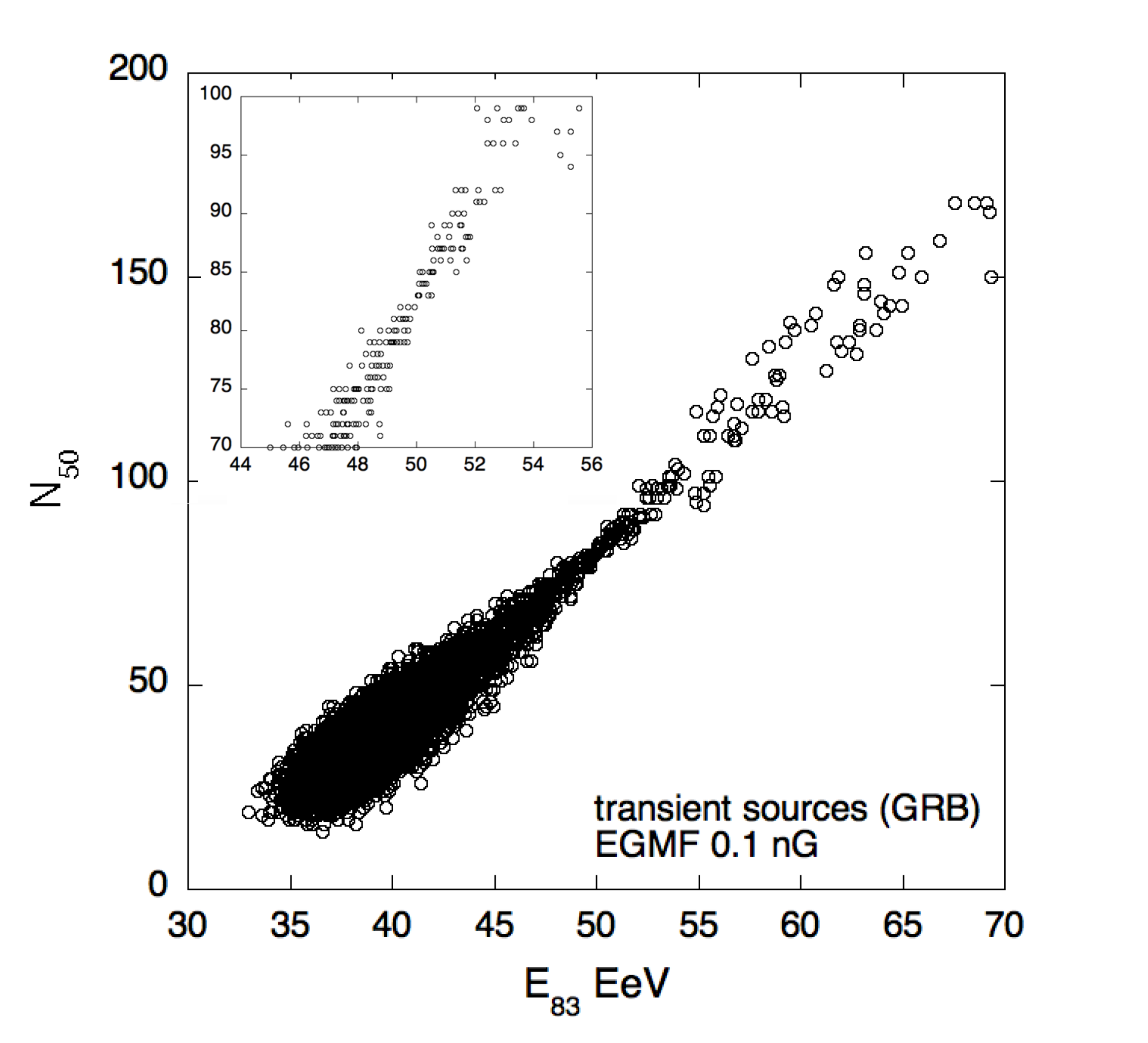}
\caption{Correlation between the number of events above 50~EeV, $N_{50}$, and the energy of the 83$^{\mathrm{rd}}$ most energetic event, $E_{83}$, in the TA-like simulated data sets, for the GRB transient source model with $B_{\mathrm{EGMF}} = 0.1$~nG.}
\label{fig:N50E83Correl}
\end{center}
\end{figure}

{\color{black}

\subsubsection{Flux analysis}

We first analyze the data sets from the point of view of their flux excess in the TA sky. After ordering the UHECRs in the simulated TA-like data sets by decreasing energies, we ask what is the energy of the 83$^{\mathrm{rd}}$ event, noted as $E_{83}$~EeV in the actual TA data set. We  then use the value of $E_{83}$ in the simulated data sets as a criterion to determine whether they may be compatible with the TA data. Both values, $N_{50}$ and $E_{83}$, are strongly correlated: the higher the energy flux, the larger the energy of the 83$^{\mathrm{rd}}$ event. This is illustrated in Fig.~\ref{fig:N50E83Correl}. 
This alternative criterion has a practical advantage in our analyses, because the sky maps we build always have the same number of events in the TA sky, namely 83, to allow for a direct quantitative comparison of their anisotropy patterns with those of the TA data, as discussed below.
}

\subsubsection{Anisotropy analysis}
\label{sec:aniso_analysis}
\label{sec:2pthotspot}

Once generated, the data sets are analysed from the point of view of their intrinsic anisotropy. For this, we use two standard analyses, which have been applied by the TA and Auger collaborations to their own data. The first one is the so-called 2-point correlation function analysis. It consists in computing the number of pairs of events with an angular separation smaller than a given angle $\theta$, and compare this number to the numbers of pairs obtained from random data sets with the same total number of events, but built from a purely isotropy flux. One determines, for each angular scale $\theta$, the fraction of isotropic data sets which have a larger number of pairs within that angular separation than the data set under study. This can be noted $\mathcal{P}_{\mathrm{iso}}(\theta)$, as it gives the probability that a purely isotropic UHECR flux would produce at least as many pairs of events separated by an angle lower than $\theta$. In the following, we concentrate on the smallest value of $\mathcal{P}_{\mathrm{iso}}(\theta)$, as $\theta$ varies from $1^{\circ}$ to $45^{\circ}$, which we note $\mathcal{P}_{\min}$. The corresponding angular scale is noted $\theta_{\min}$. {\color{black}In the actual TA data set, the value of $\mathcal{P}_{\min} \simeq 4\,10^{-4}$ is reached at the angular scale $\theta_{\min}\simeq 25^{\circ}$ \citep{Tinyakov15}.}

The second analysis of the anisotropy of the simulated data sets is a simple clustering analysis. It follows closely the analysis performed by the TA collaboration, which led to the identification of their so-called hotspot \ref{TAPaperOnHotspot}. For each simulated data set, we apply a circular ``top hat'' with an angular size of 20$^{\circ}$ around a given position in the sky, and simply count the number of events detected within that angular distance. We then compare this number with the number of events expected from a purely isotropic flux (with the same coverage map and exposure), and determine the significance of the corresponding clustering signal using the standard Li-Ma statistics \citep{LiMa83}. This significance depends on the position of the center of the circle in the sky map. After scanning the entire sky, we record the largest significance value obtained for the data set under consideration, and call it the raw (or unpenalised) hotspot significance. This value, which we note here $\sigma_{\mathrm{hotspot}}$, is then compared with the corresponding value in the actual TA dataset, $\sigma_{\mathrm{hotspot}}=5.1$.

{\color{black} While we first discuss separately each one of the two anisotropy criteria, the conjunction of both criteria must be considered when discussing the compatibility of our data sets with experimental data. For the clustering analysis, we do not perform any angular scan and only estimate the significance of the events clustering at $20^{\circ}$ as done by the TA collaboration. As a result a Li-Ma significance at $20^{\circ}$, compatible to what is observed by TA, does not guaranty by itself the compatibility of a dataset with the actual data since the clustering at lower angles could very well be much more significant. 
Likewise, fulfilling the 2-point criterion does not guarantee by itself that a significant clustering shall be detected with a top-hat analysis.}

\section{Results}
\label{sec:results}

\begin{figure}[t]
\begin{center}
\includegraphics[width=\linewidth]{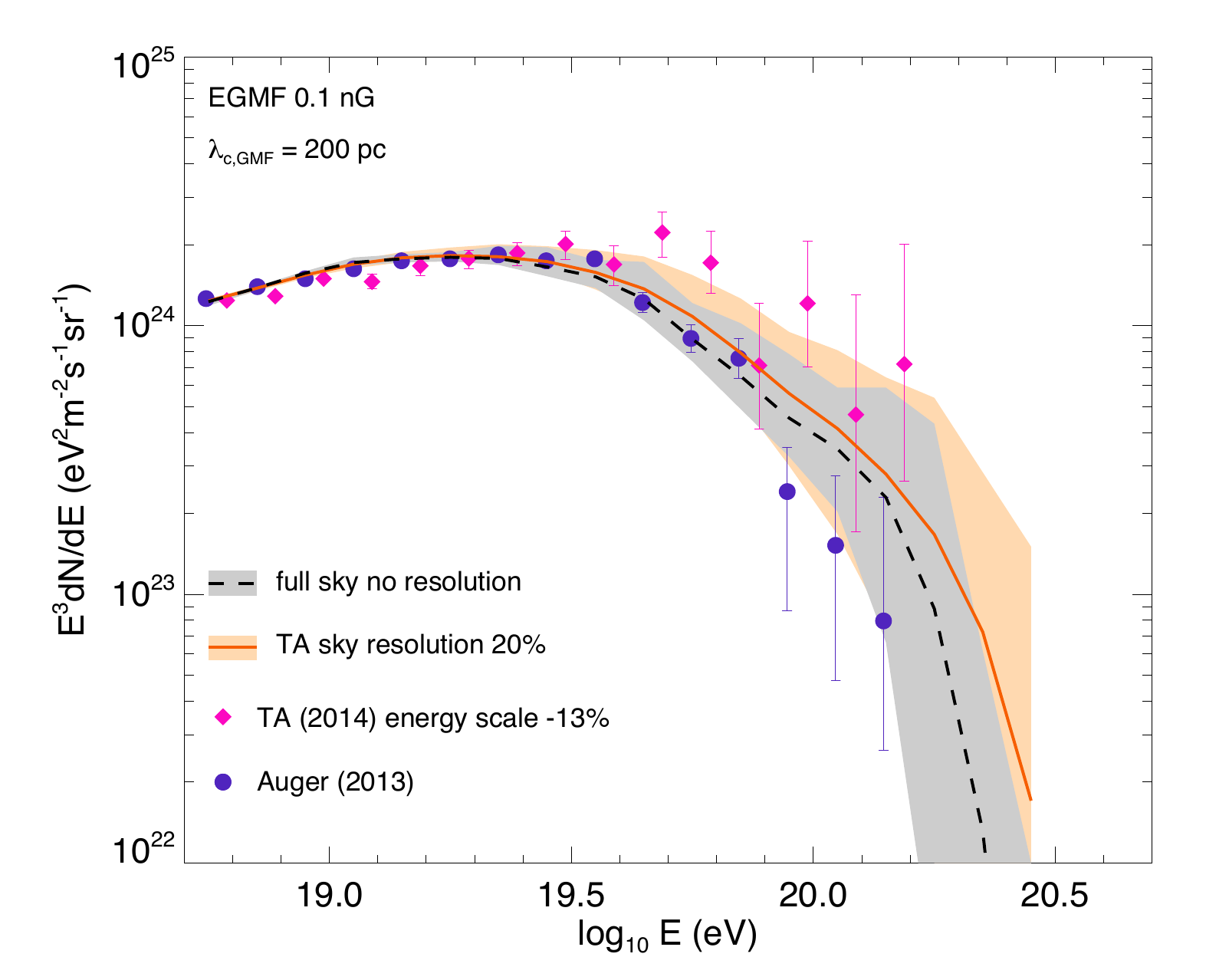}
\caption{Propagated spectra for the GRB source model with $B_{\mathrm{EGMF}} = 0.1$~nG. The dashed line shows the ``infinite statistics'' all-sky spectrum, averaged over 300 realisations of the astrophysical scenario. The gray area shows the range corresponding to the 90\% cosmic variance. The orange area corresponds to the spectra of 90\% of the 3000 TA-like simulated data sets (10 per realisation), while the orange solid line is the average of these spectra over all 3000 data sets. The Auger and TA data are also shown, with 1-$\sigma$ statistical error bars. The energy scale of the TA data set has been shifted down by 13\% \citep[see][]{Unger15}.}
\label{fig:spectrumData}
\end{center}
\end{figure}

\subsection{Transient sources}
\label{sec:transient}

We take  the GRB model \citep{Globus15a} as our transient source. This model reproduces the main features of the Auger UHECR data \citep{Globus15b}.
{\color{black} In the following, we show the results concerning the TA flux excess and anisotropy, separately, in Sect. \ref{TAfluxtr} and \ref{TAanitr}. Then we  turn to the joint analysis of the two aspects, spectrum and anisotropy 
in Sect. \ref{TAanifluxtr}.  }

\begin{figure}[t!]
\begin{center}
\includegraphics[width=\linewidth]{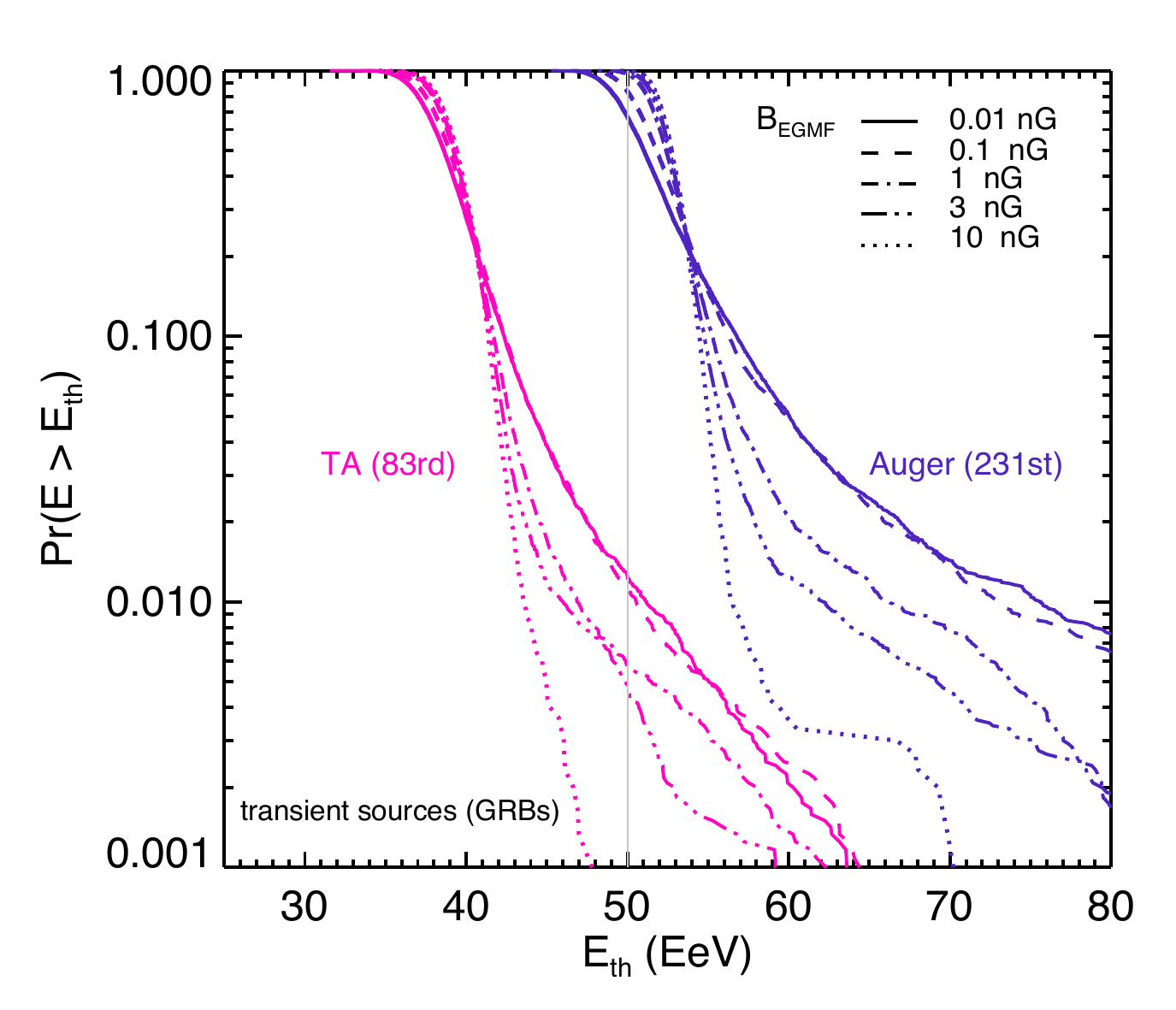}
\caption{Cumulative probability distribution of the energy of the 83$\mathrm{rd}$ highest energy events in the simulated TA-like data sets, $E_{83}$ (pink), and of the 231$^{\mathrm{st}}$ highest energy events in the simulated Auger-like data sets, $E_{231}$ (purple), for the transient source model with different values of the EGMF, as indicated. Shown is the probability that $E_{83}$ and $E_{231}$ are larger than the energy given in abscissa.}
\label{fig:Eth}
\end{center}
\end{figure}

\subsubsection{The TA flux excess}
\label{TAfluxtr}

{\color{black}Fig.~\ref{fig:spectrumData} depicts the range of variations of the energy spectra 
for $B_{\mathrm{EGMF}} = 0.1$~nG. The Auger spectrum is easily reproduced by this model. 
The TA spectrum appears as a particularly strong upward fluctuation of the average all-sky flux. This can be naturally obtained with a large contribution of a single source in the TA sky. 
The solid red line on Fig.~\ref{fig:spectrumData} shows the average spectrum assuming the TA exposure map (and an energy resolution of 20\%). As can be seen, this average spectrum is slightly higher than the average all-sky spectrum due to the TA energy resolution. The orange shaded area shows the 90\% variation range (including both cosmic variance and statistical fluctuations). The top of this area appears to be still slightly below the TA data, which indicates that a UHECR flux as high as that measured by TA might be expected is at most a few percent of the cases, in the same scenario which otherwise gives an average spectrum similar to that of Auger.}

Fig.~\ref{fig:Eth} shows the cumulative probability distribution of $E_{83}$ among all realisations of the transient source model, i.e. the probability that $E_{83}$ be larger than the energy given in abscissa, for four different values of the EGMF (pink lines).  
As expected, the value of $E_{83}$ is usually much lower than the actual value for TA, namely 50~EeV after rescaling the TA energy scale.  
In 80\% of the cases, the value of $E_{83}$ is lower than 40~EeV, whatever $B_{\mathrm{EGMF}}$. However, a small fraction of the realisations have a much larger value of $E_{83}$. These reflect the contribution of a nearby source providing a much larger number of events than average at high energy. This is more probable when the EGMF is low, because of the smaller time spread of the UHECRs reaching us from the source. The source is thus visible for a shorter amount of time, but its apparent luminosity is then correspondingly larger.

As can be seen on Fig.~\ref{fig:Eth}, for $B_{\mathrm{EGMF}}=10$~nG, it is extremely improbable that a source can have a large enough contribution in the TA sky to explain the observed excess in the UHECR flux. For $B_{\mathrm{EGMF}} = 1$~nG and a coherence length of 200~kpc (or for $B_{\mathrm{EGMF}} = 0.45$~nG and a coherence length of 1~Mpc), $E_{83}$ can reach 50~EeV in a few per mille of the realisations. Finally, for $B_{\mathrm{EGMF}} < 0.1$~nG, the occurrence rate may reach around 1\%.

On the same figure, we also plot the value of $E_{231}$ for the Auger-like simulated data sets {\color{black} (blue lines)}. The actual value of $E_{231} = 52$~EeV is quite common since the median value of the $E_{231}$ distribution varies from 51 to 53 depending on the assumed value of the EGMF.

\begin{figure}[t!]
\begin{center}
\includegraphics[width=\linewidth]{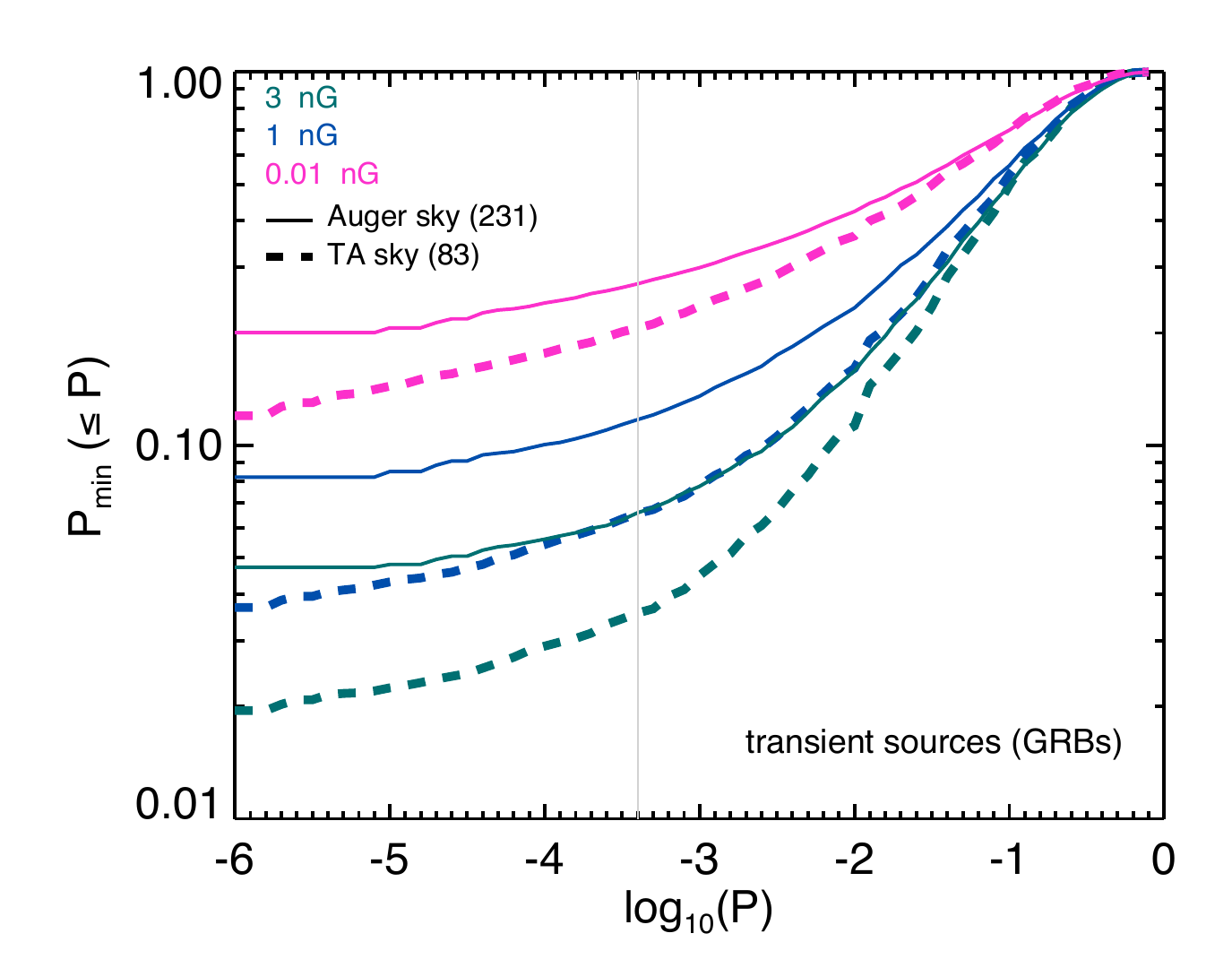}
\caption{Cumulative probability distribution of the value of $\mathcal{P}_{\min}$ associated with the 2-point correlation function analysis. All curves correspond to the case of transient sources, with the three indicated values of the magnetic field (and a coherence length $\lambda_{c} = 0.2$ Mpc). The solid (resp. dashed) lines are for Auger-like (resp. TA-like) simulated data sets, with the 231 (resp. 83) most energetic events.}
\label{fig:pminTransient}
\end{center}
\end{figure}

\begin{figure}[t!]
\begin{center}
\includegraphics[width=\linewidth]{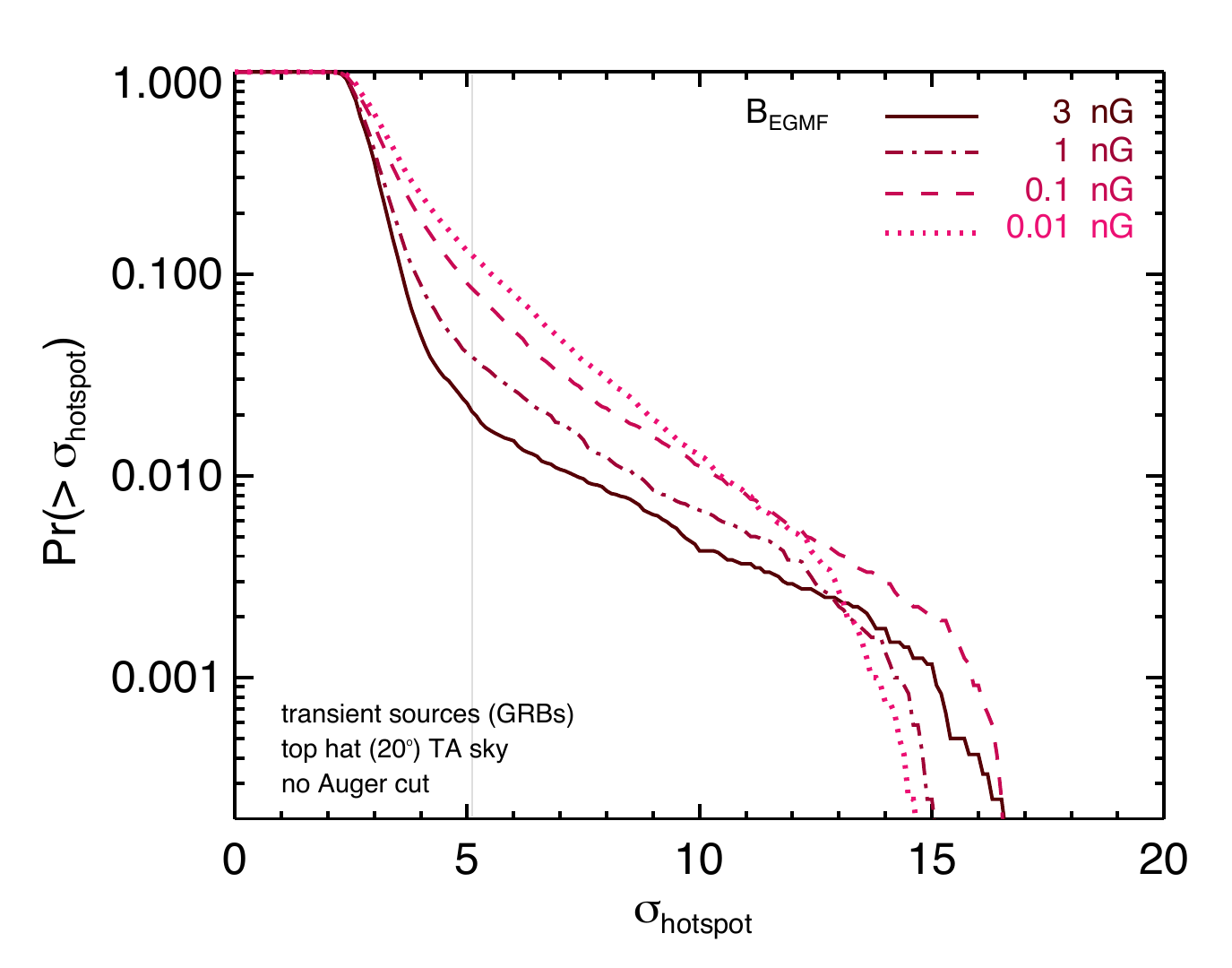}
\caption{Cumulative probability distribution of the value of $\sigma_{\mathrm{hotspot}}$ associated with the clustering analysis with a 20$^{\circ}$ top-hat filter on the TA-like data sets simulated within the transient source scenario, for different values of the EGMF, as indicated.}
\label{fig:CumulSigmaTransient}
\end{center}
\end{figure}

\subsubsection{The TA anisotropy patterns}\label{TAanitr}


Fig.~\ref{fig:pminTransient} shows the cumulative probability distribution of the value of $\mathcal{P}_{\min}$ (defined in Sect.~\ref{sec:aniso_analysis})  {\color{black} irrespective of the angular scale $\theta_{min}$ at which the minimum occurs. It allows us to compare the status of the simulated data sets with respect to the 2-point correlation function with that of the actual TA data set, for which $\mathcal{P}_{\min} \simeq 4\,10^{-4}$.}

As can be seen, such a level of anisotropy in the TA sky is rather common if the EGMF is low: about 25\% of the data sets have an anisotropy signal equal or larger to that of TA for $B_{\mathrm{EGMF}} = 0.01$~nG (with $\lambda_{c} = 0.2$ Mpc). For larger magnetic fields, the probability is smaller:  $\sim8$\% for $B_{\mathrm{EGMF}} =1$~nG, and $\sim5$\% for $B_{\mathrm{EGMF}}=3$~nG.

However, Fig.~\ref{fig:pminTransient} shows that much larger anisotropies are also possible, with occurrence rates not significantly lower than the previous ones. The probability for a data set to have $\mathcal{P}_{\min} \leq 10^{-6}$ 
is only $\sim 2$ times lower than the probability to have $\mathcal{P}_{\min} \leq 4\,10^{-4}$. In other words, a large fraction of the anisotropic data sets are in fact \emph{too} anisotropic to be compatible with the TA data, and must be rejected. {\color{black} A similar conclusion can be reached from the study of clustering in the data sets, (see Sect.~\ref{sec:2pthotspot}, and   Fig.~\ref{fig:CumulSigmaTransient}.)}

{\color{black} 

The data sets with a too low value of $\mathcal{P}_{\min}$ or a too large value of $\sigma_{\mathrm{hotspot}}$, are expected to be typically those showing a very large contribution of one individual source. This raises an important and general problem: while
flux excesses similar to that of TA can be obtained in $\sim$ a percent of the realisations for low values of the EGMF (see above), it is very likely that these particular realizations are those which give rise to too large an anisotropy. To investigate this, we now turn to the joint analysis of the two aspects, spectrum and anisotropy, of the simulated data sets.
}

\begin{figure}[t!]
\begin{center}
\includegraphics[width=\linewidth]{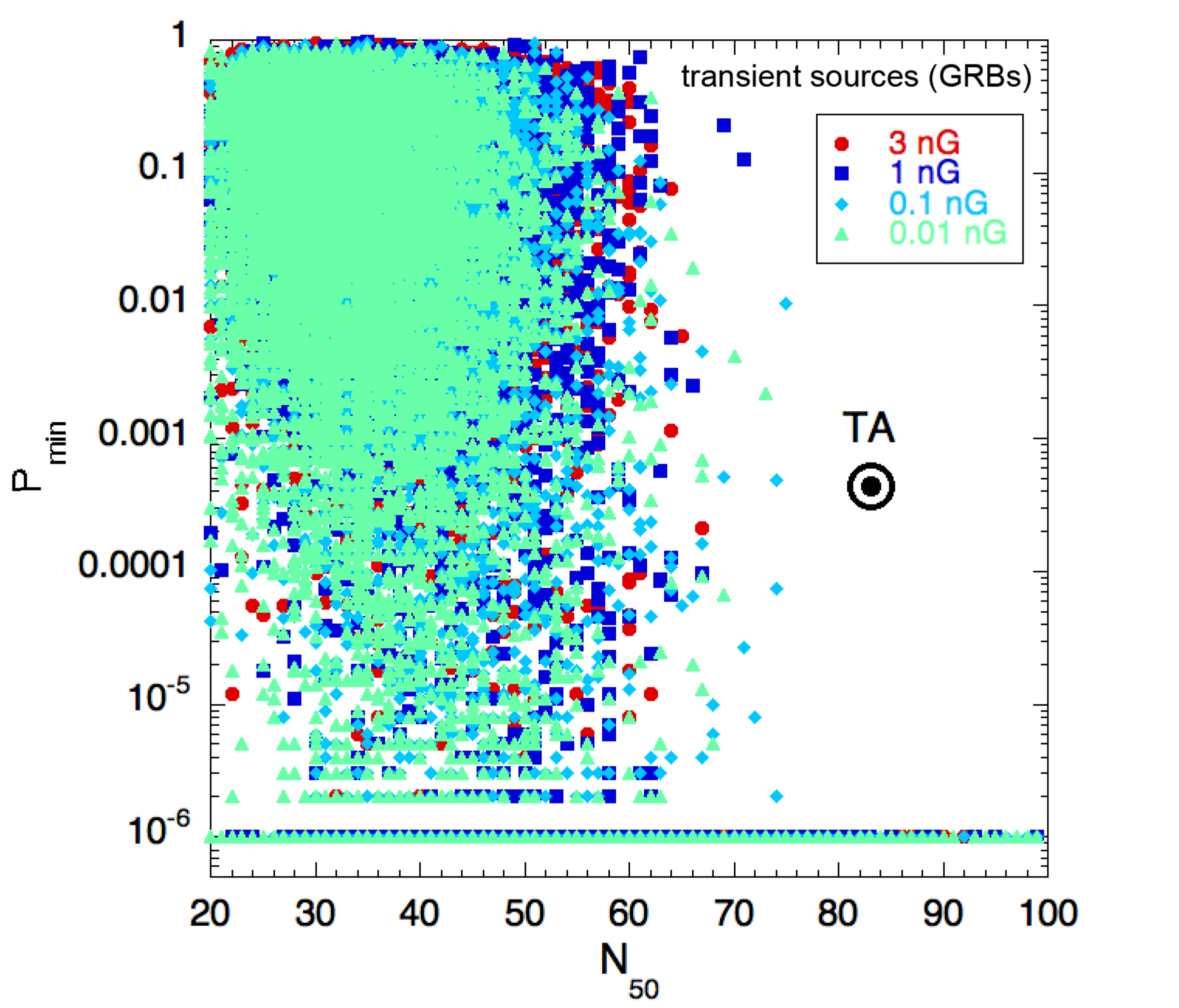}
\caption{A scatter plot of the values of $\mathcal{P}_{\min}$ vs. $N_{50}$ for all the TA-like data sets simulated in the transient source scenarios, with 4 values of the EGMF, as indicated. The position of the actual TA data set is indicated by the $\odot$ symbol. 
Note that only the realisations fulfilling both the Auger 2-point and Auger flux criteria  are considered in this scatter plot. }
\label{fig:Correl2ptN50Transient}
\end{center}
\end{figure}

\subsubsection{A joined analysis of the spectrum and anisotropy constraints}\label{TAanifluxtr}

{\color{black}Before turning to the TA data we note that from now on, we apply a selection criterion to the Auger-like simulated data set. According to this criteria  we reject realisations for which the flux excess in the TA sky is due to a dominant source which also contributes significantly to the UHECR flux in the Auger sky, thereby producing a spectrum that does not match the Auger spectrum {\color{black} and/or produces a strong anisotropy} at high energy.
Since our Auger-like sky maps are built with 231 events,  
we use the value of the energy of the $231^{\mathrm{st}}$ event, $E_{231}$, to build a selection criterion. 
Specifically, we reject all realisations for which 7 or more of its 10 data sets have $E_{231} \geq 57$~EeV. This flux excess above 52 EeV of about 20-25\% (compared to the Auger data), corresponds to a significance of $\sim 3 \sigma$. Likewise from 
we require that the 231~highest energy events in the Auger sky do not show a much stronger anisotropy than the actual Auger data. 
We use again $\mathcal{P}_{\min}$ as the relevant quantity, and reject all realisations for which 7 or more of the 10 data sets have a $\mathcal{P}_{\min}$ value smaller than $4\,10^{-4}$. In the following we refer to these criteria as the "Auger flux" and  the "Auger 2-point" cuts.}
{\color{black}  The combined effect of these two cuts is to reject between 10 and 30\% of the total number of realisations depending on the astrophysical model (as can be easily deduced from Figs.~\ref{fig:Eth} and \ref{fig:pminTransient} for the transient scenario).}

 Fig.~\ref{fig:Correl2ptN50Transient} shows the number of events above 50~EeV and the maximal departure from isotropy described by the 2-point correlation function.  In this scatter plot, each dot corresponds to a TA-like simulated data set 
with $N_{50}$ in abscissa and $\mathcal{P}_{\min}$ in ordinate {\color{black}(again irrespective of the value of $\theta_{\min}$)}, for 4 different values of the EGMF, 
all with a coherence length of 0.2~Mpc (see Sect.~\ref{sec:GMFEGMF} for scaling). The position of the TA data set ($N_{50} = 83$ and $\mathcal{P}_{\min}=4\,10^{-4}$) in this $\mathcal{P}_{\min}$--$N_{50}$ plane is shown by the $\odot$ symbol. 

Not surprisingly, most simulated data sets have much lower values of $N_{50}$, around 30--45, as would be the case in the actual TA data set if the TA spectrum were similar to that of Auger. It can also be seen that data sets with an acceptable value of $\mathcal{P}_{\min}$ are not uncommon. As a matter of fact, Fig.~\ref{fig:pminTransient} corresponds to the projection of this scatter plot on the y-axis ({\color{black} after accounting for the "Auger flux" and the "Auger 2-point" cuts}) \footnote{In order to evaluate $\mathcal{P}_{\min}$, we computed one million random realisations of an isotropic flux. Therefore, we cannot attribute values to $\mathcal{P}_{\min}$ lower than $10^{-6}$. {\color{black}  The data sets corresponding to the dots on the 
$\mathcal{P}_{\min}=10^{-6}$ line are thus in fact  data sets which have $\mathcal{P}_{\min}\leq10^{-6}$.} }
Data sets for which the value of $N_{50}$ is large enough to be compatible with the TA spectrum do exist: they correspond to the small fraction of data sets with large values of $E_{83}$ in Fig.~\ref{fig:Eth}. {\color{black} However, almost all of them have prohibitive values of $\mathcal{P}_{\min}$, due to the concentration of events on small angular scales from the very bright source, which is responsible for the flux excess.} {\color{black}As a result, all our data sets with $N_{50}>74$ turn out to show values of $\mathcal{P}_{\min}\leq 10^{-6}$ as anticipated in Sect.~\ref{TAanitr}.}

\begin{figure}[t!]
\begin{center}
\includegraphics[width=\linewidth]{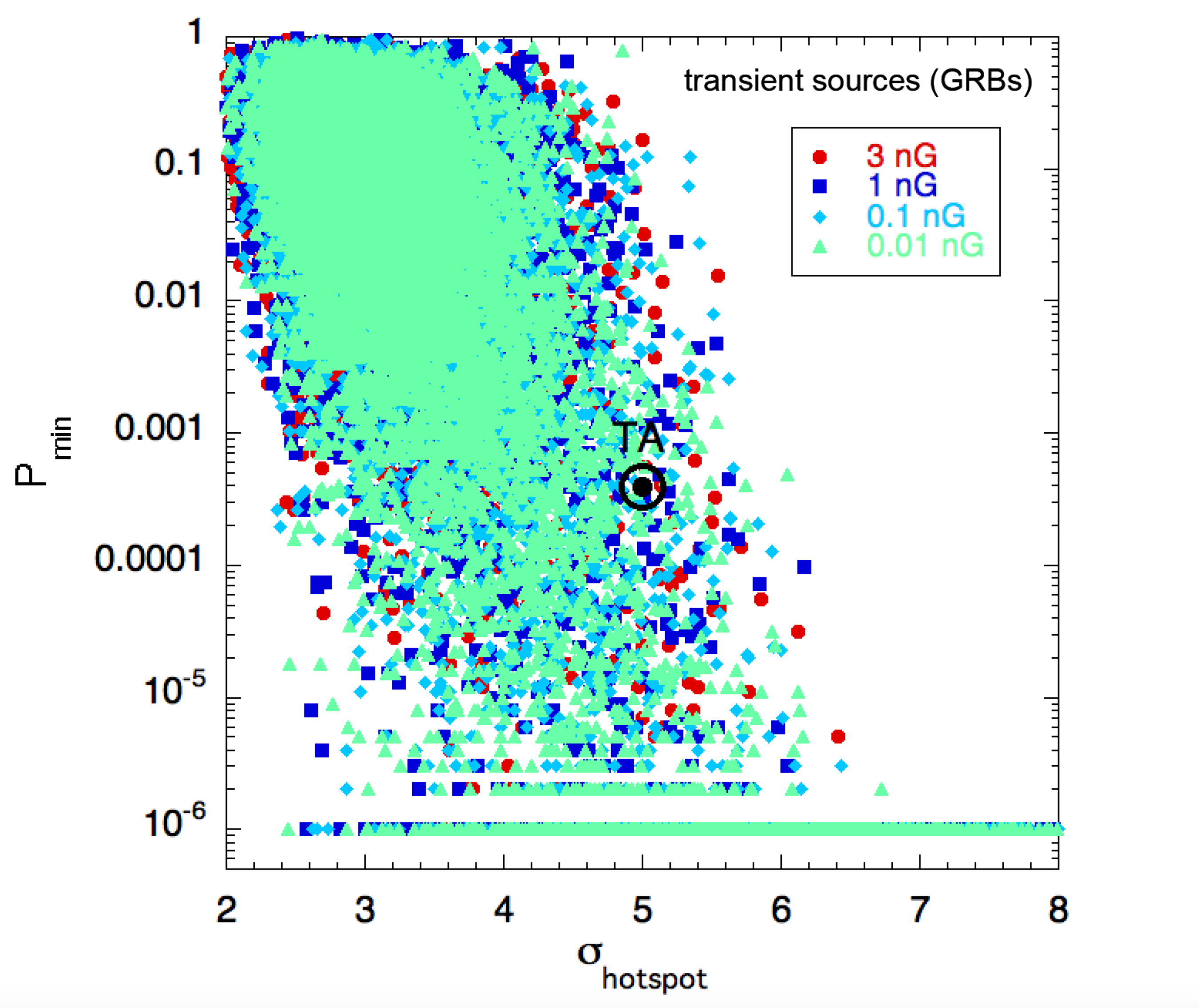}
\caption{Same as Fig.~\ref{fig:Correl2ptN50Transient}, with the values of $\mathcal{P}_{\min}$ vs. $\sigma_{\mathrm{hotspot}}$.}
\label{fig:PminNsigmaTransient}
\end{center}
\end{figure}
{\color{black}
Fig.~\ref{fig:PminNsigmaTransient}, depicts a scatter plote of the simulated values in the   $\mathcal{P}_{\min}$ and $\sigma_{\mathrm{hotspot}}$ plane. Although most data sets have a value of $\sigma_{\mathrm{hotspot}}$ between 2 and 4, a sizable fraction of the data sets with an appropriate value of $\mathcal{P}_{\min}$ shows clustering properties compatible with the TA data, with $\sigma_{\mathrm{hotspot}}$ around 5, rather independently of the EGMF value. }
{\color{black}Note that while $\sigma_{\mathrm{hotspot}}$ and $\mathcal{P}_{\min}$ are clearly correlated, the two anisotropy observables are not redundant: appropriate values of $\sigma_{\mathrm{hotspot}}$ can be associated with values of $\mathcal{P}_{\min}$ much lower than what is actually observed, all the more when the significance of the event clustering is maximal at angular scales much lower than $20^\circ$.}

\begin{figure}[t!]
\begin{center}
\includegraphics[width=\linewidth]{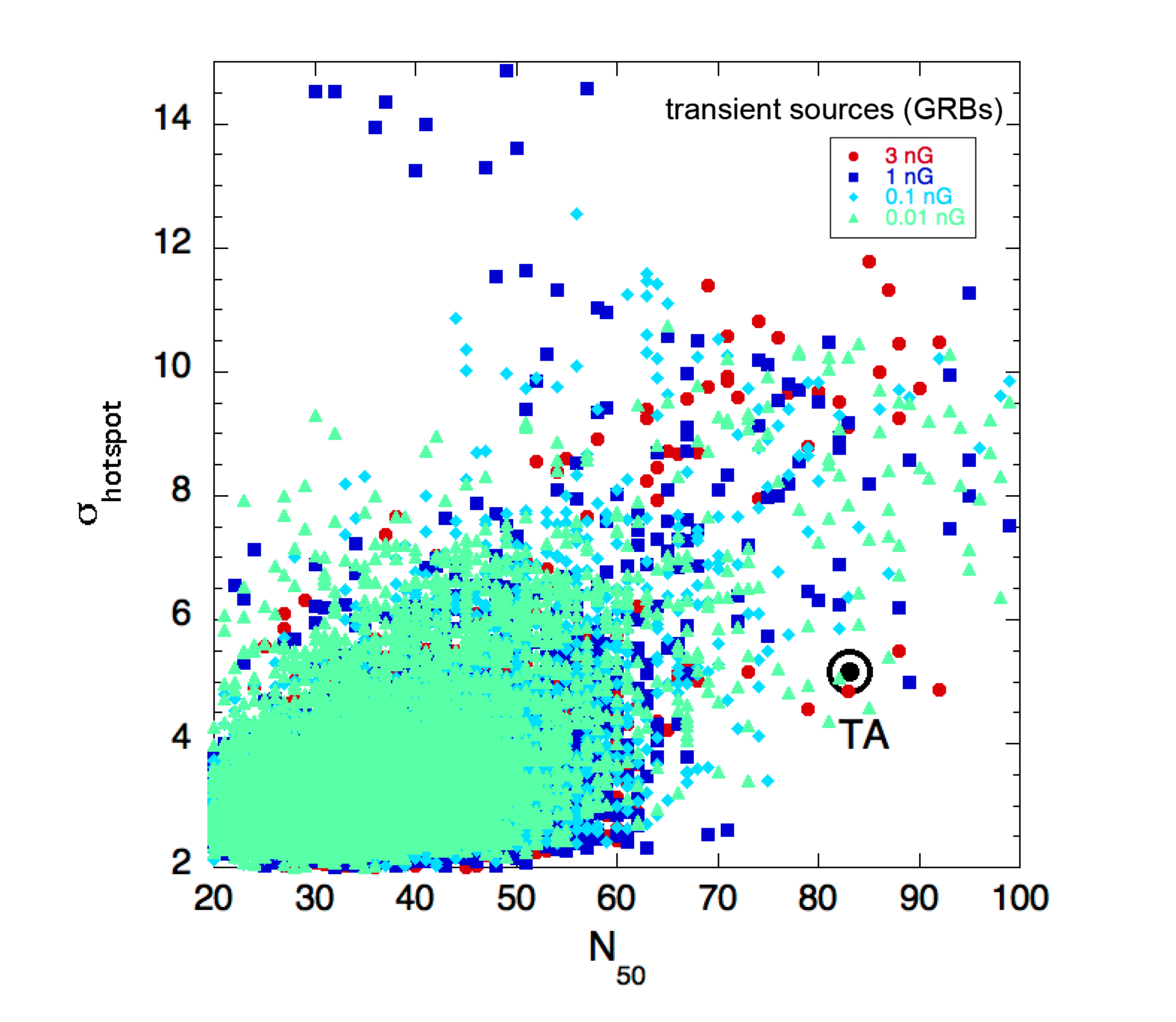}
\caption{Same as Fig.~\ref{fig:Correl2ptN50Transient}, with the values of $\sigma_{\mathrm{hotspot}}$ vs. $N_{50}$.}
\label{fig:correlSigmaN50}
\end{center}
\end{figure}

{\color{black}  Fig.~\ref{fig:correlSigmaN50} compares the clustering properties at $20^\circ$,  $\sigma_{\mathrm{hotspot}}$, with the flux properties,  $N_{50}$. {\color{black}Values of $\sigma_{\mathrm{hotspot}}$ above 7 are most common for data sets showing values of $N_{50}$ close to that observed by TA. A small fraction of these data sets, however shows  hotspot significances closer to the TA observation. As it turns out, however, all of those have much more significant clustering at lower angular scales, and their 2-point correlation functions are incompatible with the actual data (see Fig.~\ref{fig:Correl2ptN50Transient}).}

}

\begin{figure*}[ht!]
\begin{center}
\includegraphics[width=0.5\linewidth]{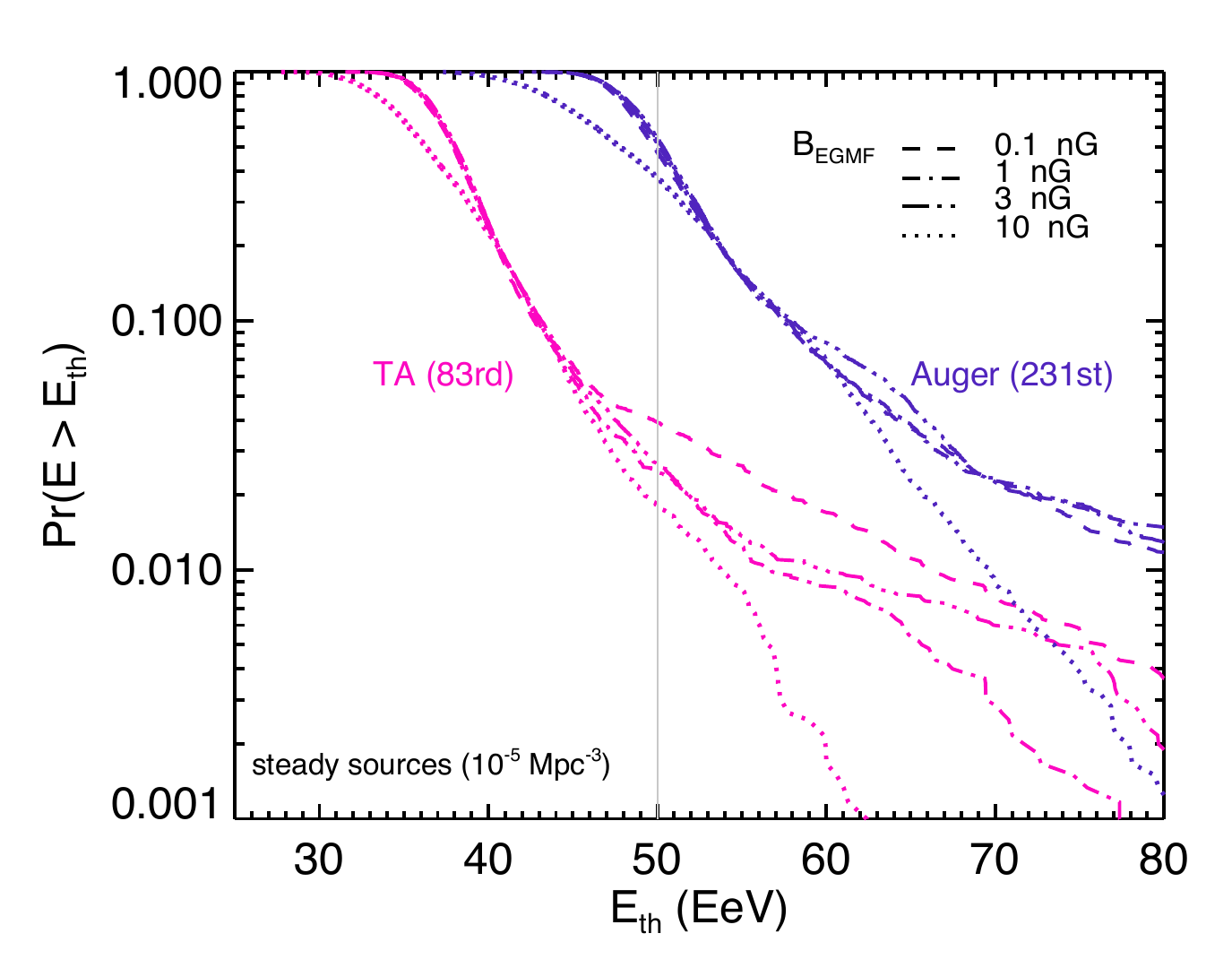}\includegraphics[width=0.5\linewidth]{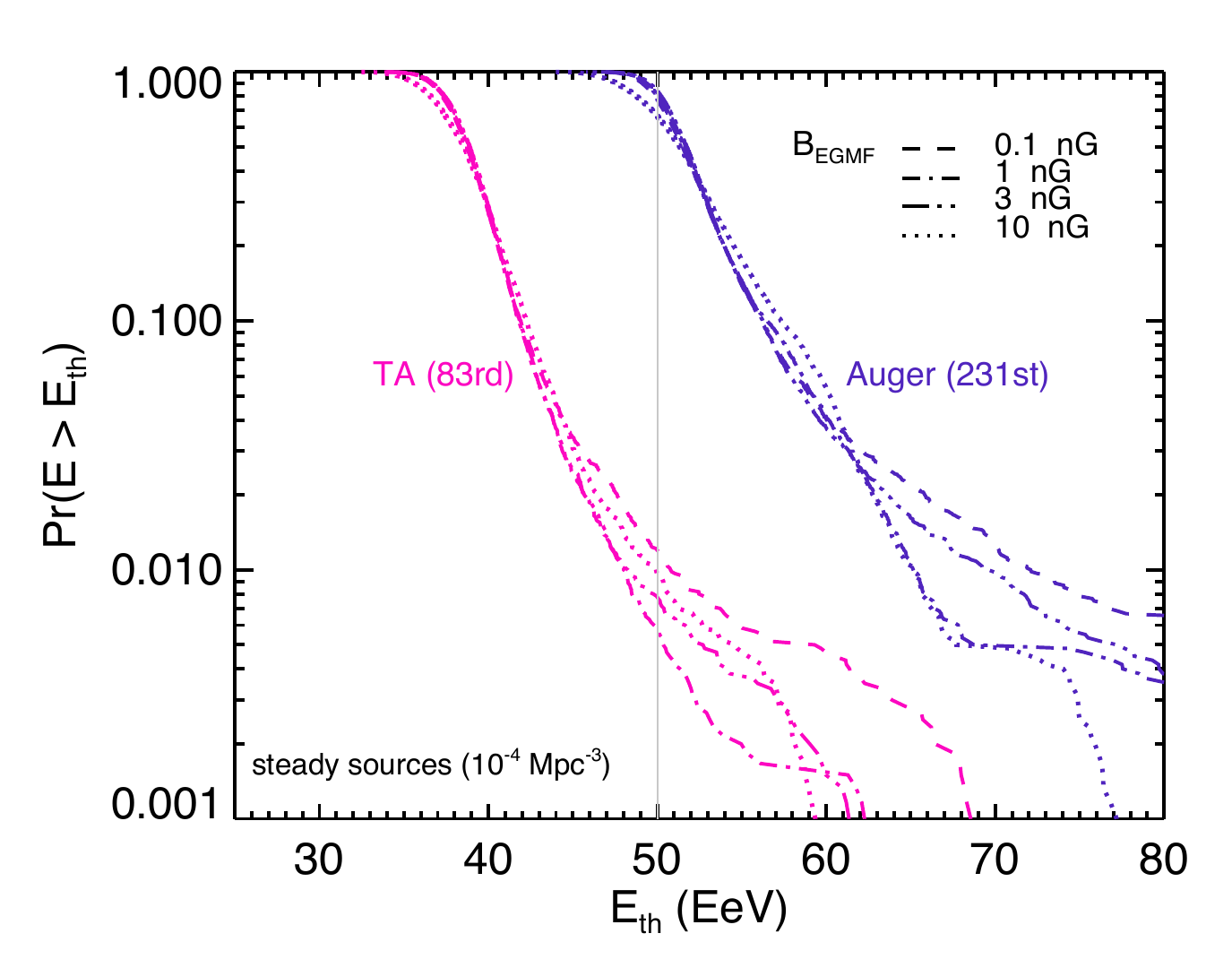}
\caption{
The cumulative probability distribution of the energy of the 83$\mathrm{rd}$ highest energy events in the simulated TA-like data sets, $E_{83}$ (pink), and of the 231$^{\mathrm{st}}$ highest energy events in the simulated Auger-like data sets, $E_{231}$ (purple), for the steady source model with different values of the EGMF, as indicated. The left and right plots correspond to  source densities of $10^{-5}\,\mathrm{Mpc}^{-3}$ and  $10^{-4}\,\mathrm{Mpc}^{-3}$ respectively.}
\label{fig:EthSteady}
\end{center}
\end{figure*}

We conclude that the transient source scenario does not appear to be a likely candidate to account for the currently available observational data, at least under the general assumptions inherent to our model. 


\begin{figure*}[t!]
\begin{center}
\includegraphics[width=0.5\linewidth]{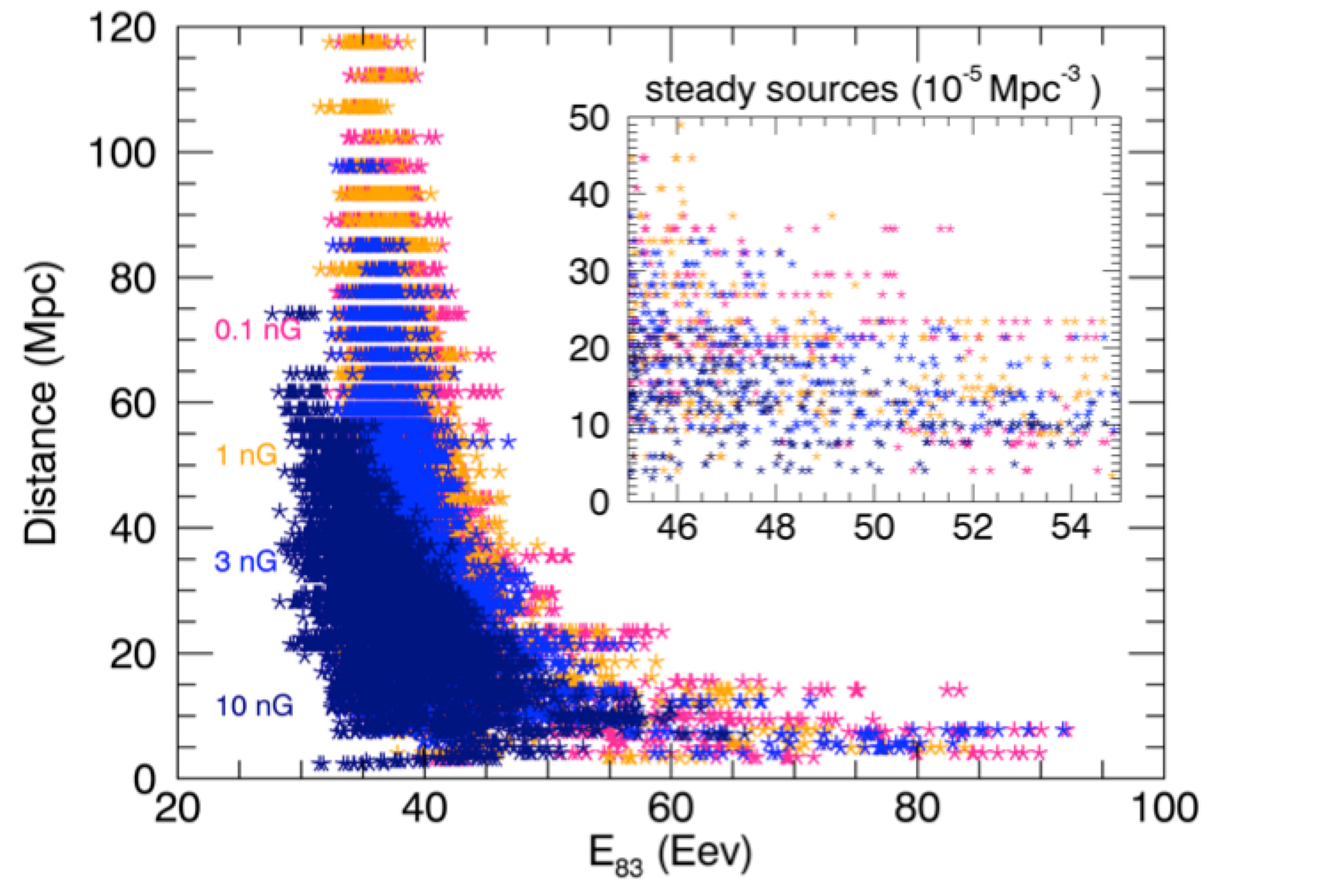}\includegraphics[width=0.5\linewidth]{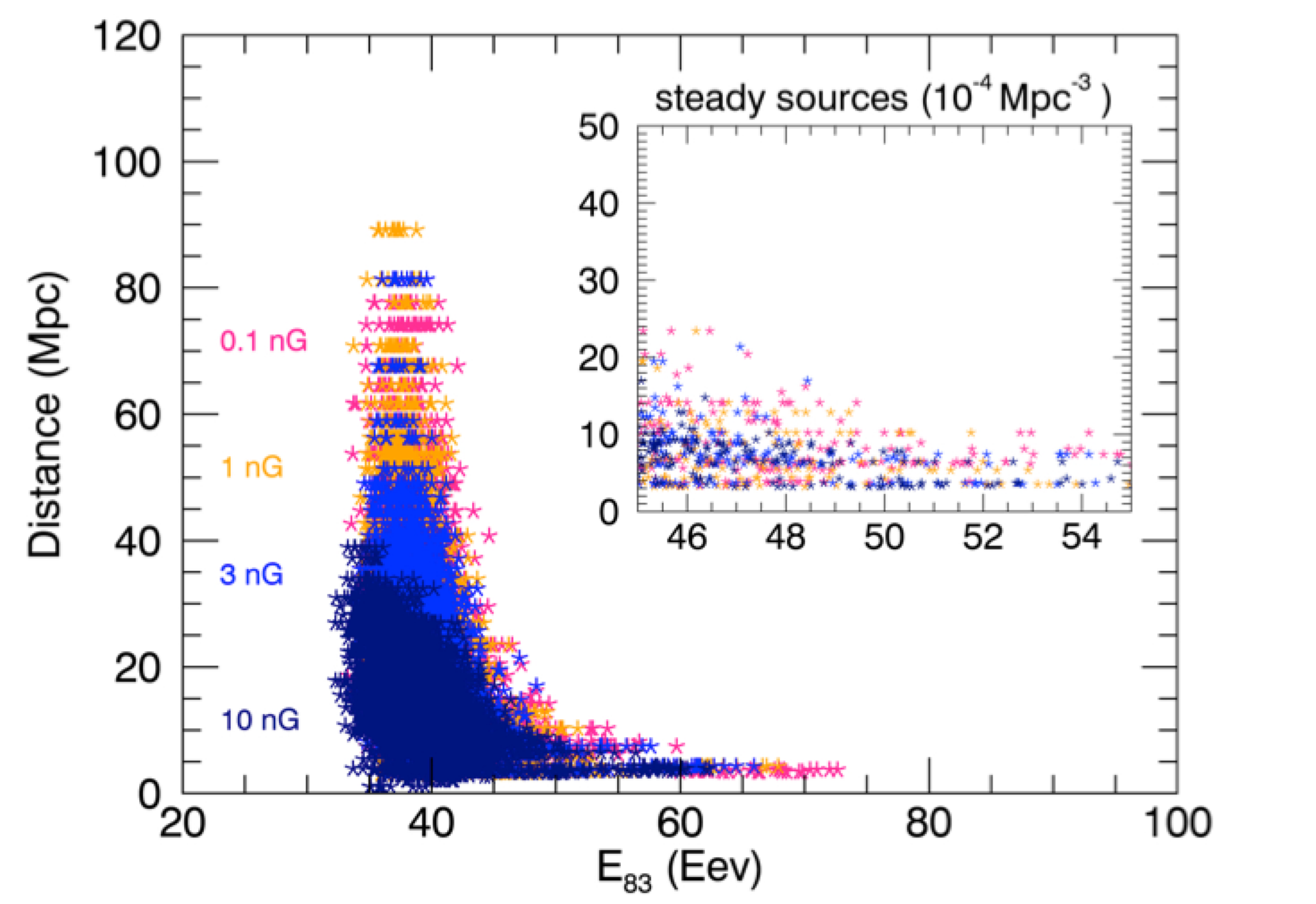}
\caption{A scatter plot of the values of the source distance vs. $E_{83}$ for all the TA-like data sets simulated in the steady source scenario, for a source density of $10^{-5}$ Mpc$^{-3}$ (left) and $10^{-4}$ Mpc$^{-3}$ (right), with 4 values of the magnetic field, as indicated. The insets show a zoom on the  $47 \le E_{83} \rm (EeV)  \le 53$ range close to the TA data. In that case the source distance is $\lesssim 20$ Mpc for a source density of $10^{-5}$ Mpc$^{-3}$ and $\lesssim10$ Mpc for a source density of $10^{-4}$ Mpc$^{-3}$.}
\label{fig:distances}
\end{center}
\end{figure*}

\subsection{Steady sources}
\label{sec:steady}

We turn  now to  steady sources and present the results of  simulations of this scenario for 4 values of the magnetic field, $B_{\mathrm{EGMF}} = 0.1$, 1, 3 and 10~$\mathrm{nG}$, and two different source densities: $n_{\mathrm{s}} = 10^{-4}$~and $10^{-5}\,\mathrm{Mpc}^{-3}$. Even though these are obviously not GRBs we use as a source model the spectrum, composition and luminosity function of the GRB model \citep{Globus15a} that fits the observed Auger data. 
For steady sources, the propagation time delay of the particles and its spread have no direct incidence on the apparent luminosity of the source. However, the Galactic and extragalactic magnetic fields play a role in spreading the UHECRs.

\subsubsection{The TA flux excess}


{\color{black} 
The fractions of simulated data sets with $E_{83}$ and $E_{231}$ larger than a given energy are shown in Fig.~\ref{fig:EthSteady}. The left and right panels correspond  to source densities of $n_{\mathrm{s}} = 10^{-5}\,\mathrm{Mpc}^{-3}$ and $10^{-4}\,\mathrm{Mpc}^{-3}$, respectively. 
Values of $E_{83}$ in the TA sky larger than 50~EeV, corresponding to an excess of the  flux similar to that measured by TA, are relatively rare. 
For $n_{\mathrm{s}} = 10^{-5}\,\mathrm{Mpc}^{-3}$, around 2\% to 4\% of the data sets satisfy the criterion, while this fraction drops to around 1\% for $n_{\mathrm{s}} = 10^{-4}\,\mathrm{Mpc}^{-3}$. This fraction does not depend much on the EGMF value, contrary to the transient source case where larger fields resulted in larger time spreads for the particles and thus weaker flux excesses. The effect of the EGMF is nevertheless visible for the lower source density, where the lower probabilities obtained for large magnetic fields (and also the lower value of the median of the $E_{231}$ distribution) 
are due to the magnetic horizon effect, which attenuates the flux of UHECRs received from sources at intermediate distances.}

Figure \ref{fig:distances} depicts the two scatter plots corresponding to the correlation between the source distance and $E_{83}$ for the two source densities. We see that  to obtain $47\leq E_{83}{\rm (EeV)} \leq 53$ , the distance of the dominant source has to be $D\lesssim 20$ Mpc for a source density of $10^{-5}$ Mpc$^{-3}$ and $D\lesssim 10$ Mpc for a source density of $10^{-4}$ Mpc$^{-3}$.

\begin{figure*}[t!]
\begin{center}
\includegraphics[width=0.45\linewidth]{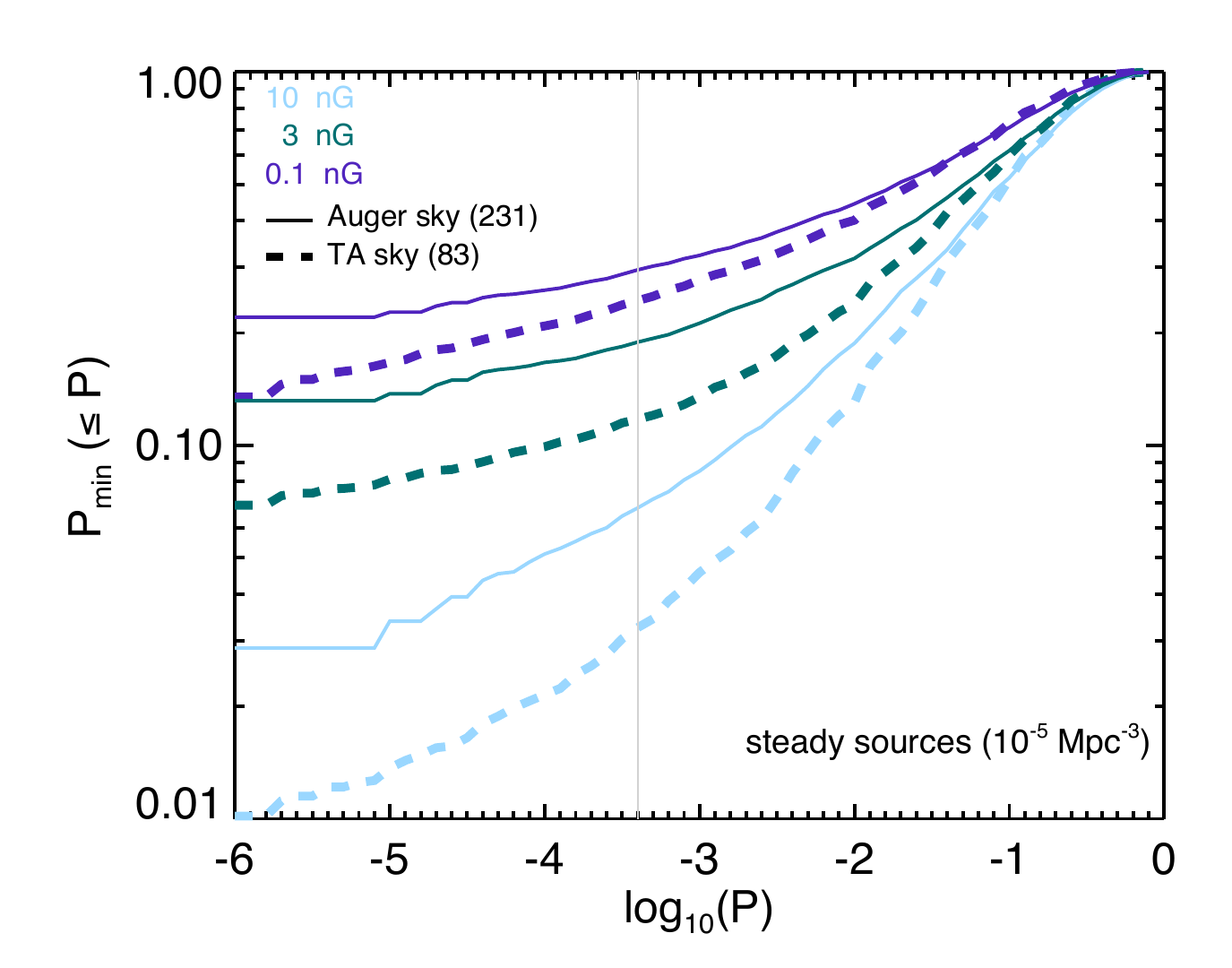}\includegraphics[width=0.45\linewidth]{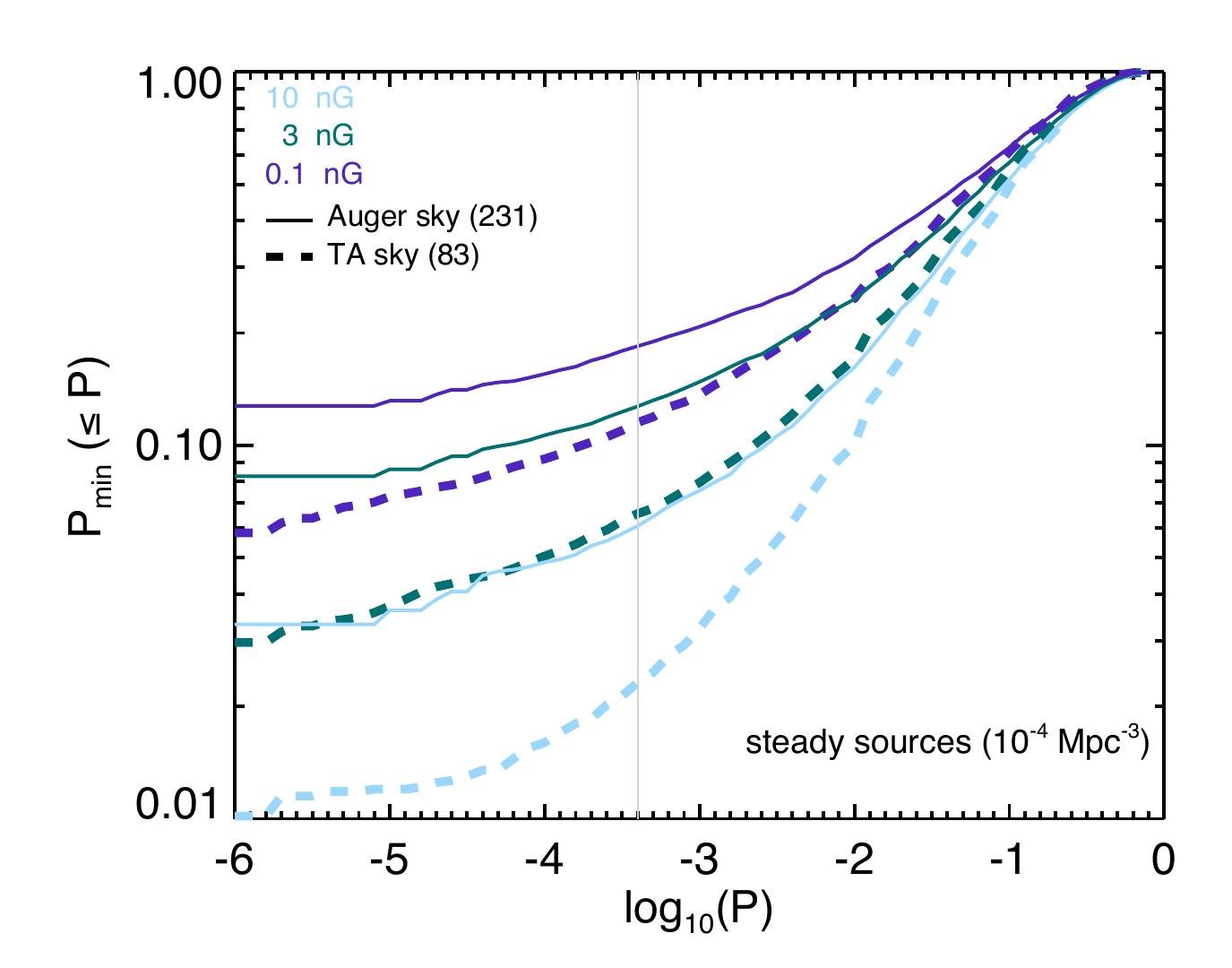}
\caption{Cumulative probability distribution of the value of $\mathcal{P}_{\min}$ associated with the 2-point correlation function analysis. All curves correspond to the case of steady sources, with the three indicated values of the magnetic field (and a coherence length $\lambda_{c} = 0.2$ Mpc). The solid (resp. dashed) lines are for Auger-like (resp. TA-like) simulated data sets, with the 231 (resp. 83) most energetic events. The left and right plots correspond to a source density of $10^{-5}\,\mathrm{Mpc}^{-3}$ and  $10^{-4}\,\mathrm{Mpc}^{-3}$ respectively.}
\label{fig:pminSteady}
\end{center}
\end{figure*}

\subsubsection{The TA anisotropy patterns}


{\color{black} We first consider the 2-point correlation function of the simulated data sets, and computed the value of $\mathcal{P}_{\min}$. Fig.~\ref{fig:pminSteady} shows the probability distribution of $\mathcal{P}_{\min}$. }
The model with a lower source density is more likely to exhibit large anisotropies. This is expected considering that fewer sources  contribute to the observed flux,  and a dominant source is more likely to contribute a large fraction of the UHECRs. Also expected is the influence of the magnetic field: larger values of the EGMF result in rarer occurrences of large anisotropies. For a source density $n_{\mathrm{s}} = 10^{-5}\,\mathrm{Mpc}^{-3}$, around 12\% of the TA-like data sets have a value of $\mathcal{P}_{\min}$ lower (and are thus more anisotropic) than the TA data set in the case of a 3 nG field, while this fraction rises to 25\% for a 0.1 nG field. These fractions are of the order of 8\% and 12\%, respectively, if the source density is $n_{\mathrm{s}} = 10^{-4}\,\mathrm{Mpc}^{-3}$.

\begin{figure}[t!]
\begin{center}
\includegraphics[width=0.9\linewidth]{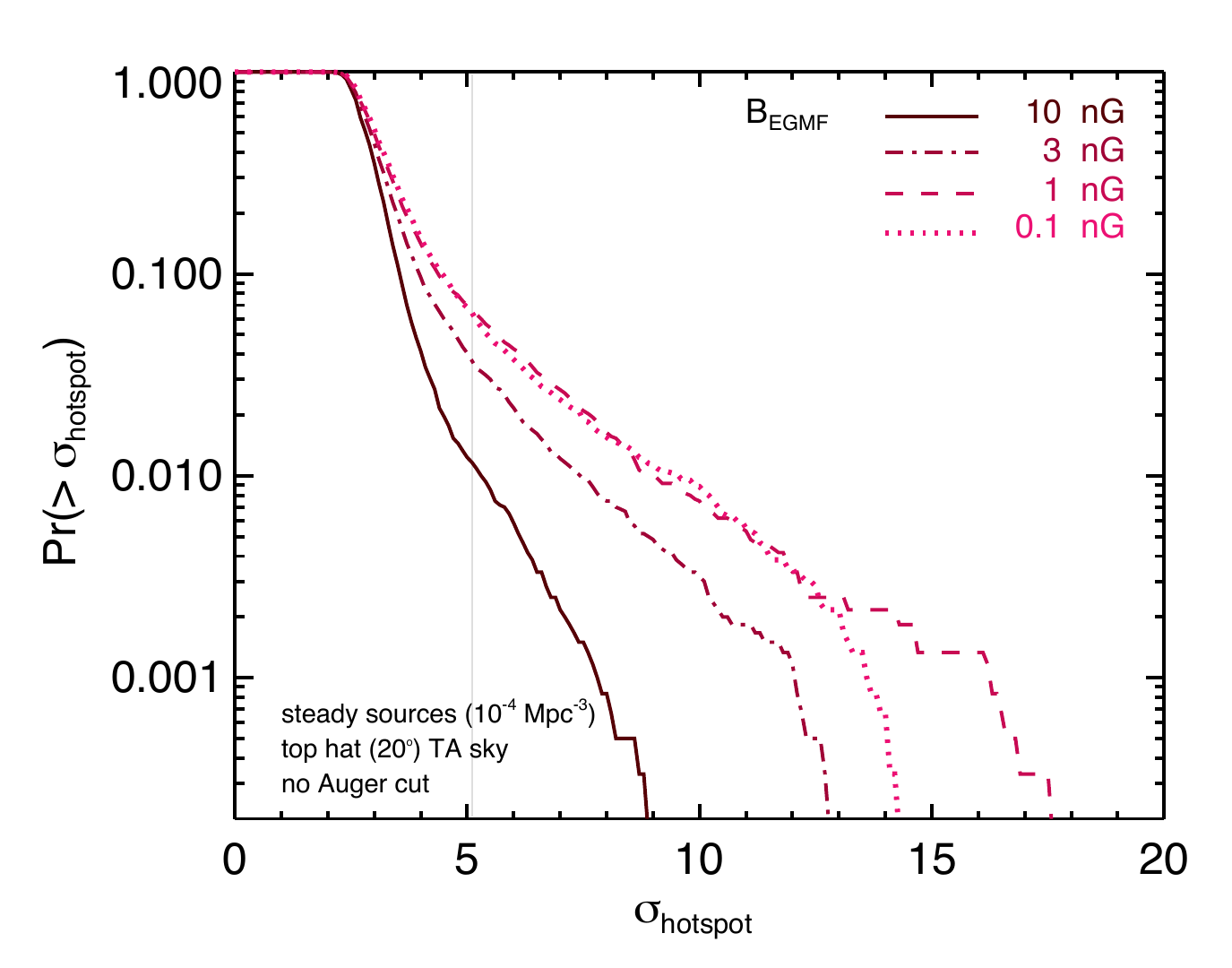}
\caption{Cumulative probability distribution of the value of $\sigma_{\mathrm{hotspot}}$ associated with the clustering analysis with a 20$^{\circ}$ top-hat filter on the TA-like data sets simulated within the steady source scenario and density $10^{-4}\,\rm{Mpc}^{-3}$, for different values of the EGMF, as indicated.}
\label{fig:CumulSigmaSteady}
\end{center}
\end{figure}

Although these numbers are reasonably large, Fig.~\ref{fig:pminSteady} shows that much larger anisotropies are also fairly common. For these two values of the EGMF between 3\% and 12\% of the TA-like simulated data sets have a value of $\mathcal{P}_{\min}$ as low as $10^{-6}$, depending on the source density. 

The cumulative distribution of the second anisotropy criterion, $\sigma_{\rm hotspot}$, 
 is shown in Fig.~\ref{fig:CumulSigmaSteady} for 4 different values of the EGMF and a density of $10^{-4}\,\rm{Mpc}^{-3}$. 
As could be anticipated, the data sets corresponding to larger values of the EGMF 
{\color{black} are less likely to show a very large anisotropy. Note that the influence of the source density is relatively moderated for this observable as expected  from our simplified analytical study (see Eq.~\ref{eq:angularSizeSteady})}. {\color{black} Again, for the low and moderate EGMF cases, even though a few percent of the data sets have clustering properties compatible with that of the TA data, a substantial fraction of those have clusters which are in fact much more significant than the TA hotspot. Now these are the most likely to be the ones with a high flux excess above 50 EeV, as required by the TA spectrum. }

{\color{black} The situation in the case of a 10 nG EGMF is quite different, as can be seen in Fig.~\ref{fig:pminSteady} and Fig.~\ref{fig:CumulSigmaSteady}. The probability of observing a strong anisotropy decreases quite drastically when passing from a 3 nG to a 10 nG EGMF. With such a large field, one  enters a regime where the isotropization time of the highest energy nuclei becomes significantly shorter than their (time-) horizon for energy losses (which is of the order of 100 Mpc/c for a 50 EeV O nucleus). For such a strong value of the EGMF, only UHECR nuclei coming from relatively nearby sources have a chance to reach  the observer without being isotropized on their way. This  results in a much lower probability to obtain a strongly anisotropic sky.  }

\begin{figure*}[t!]
\begin{center}
\includegraphics[width=0.45\linewidth]{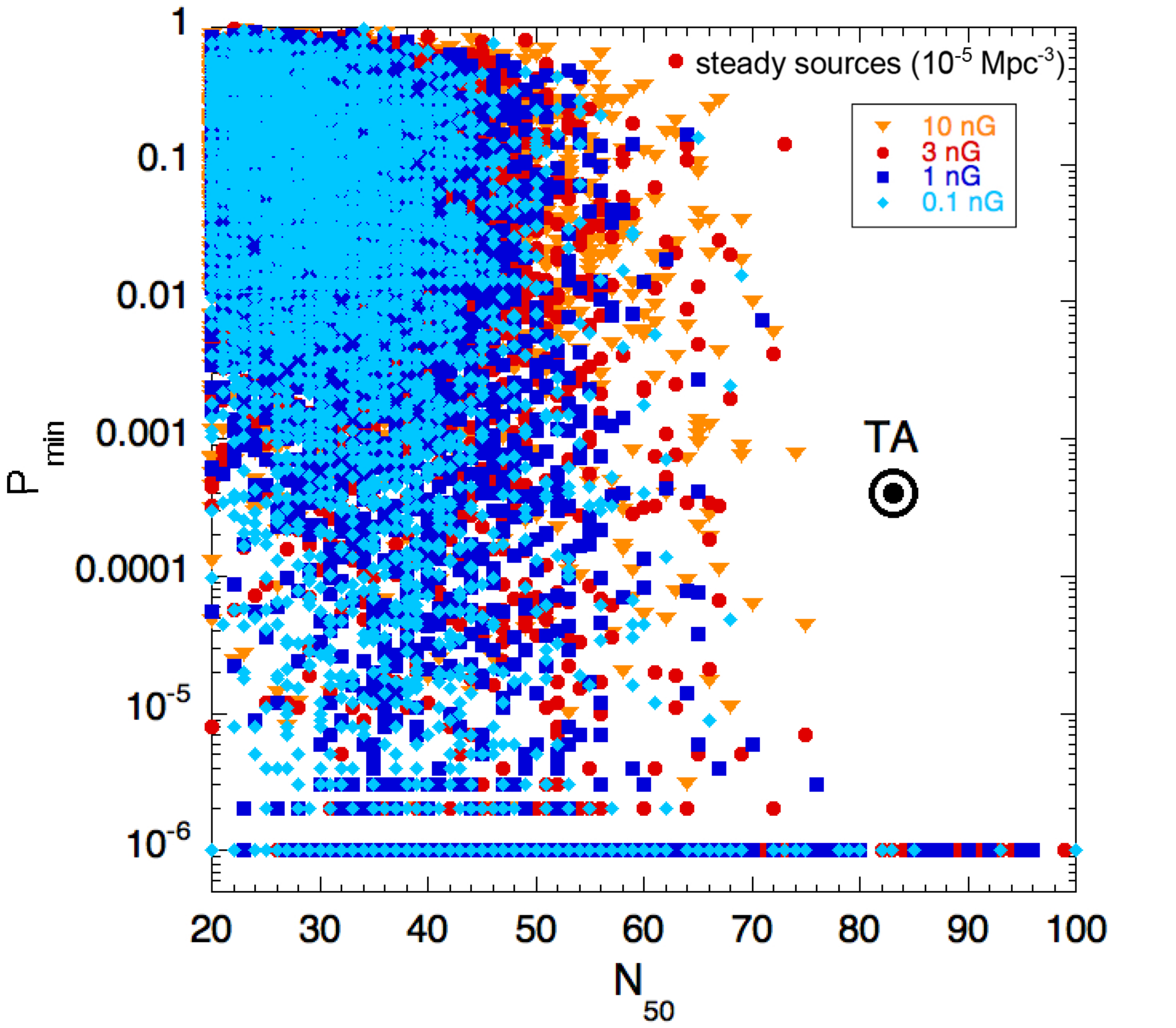}\includegraphics[width=0.45\linewidth]{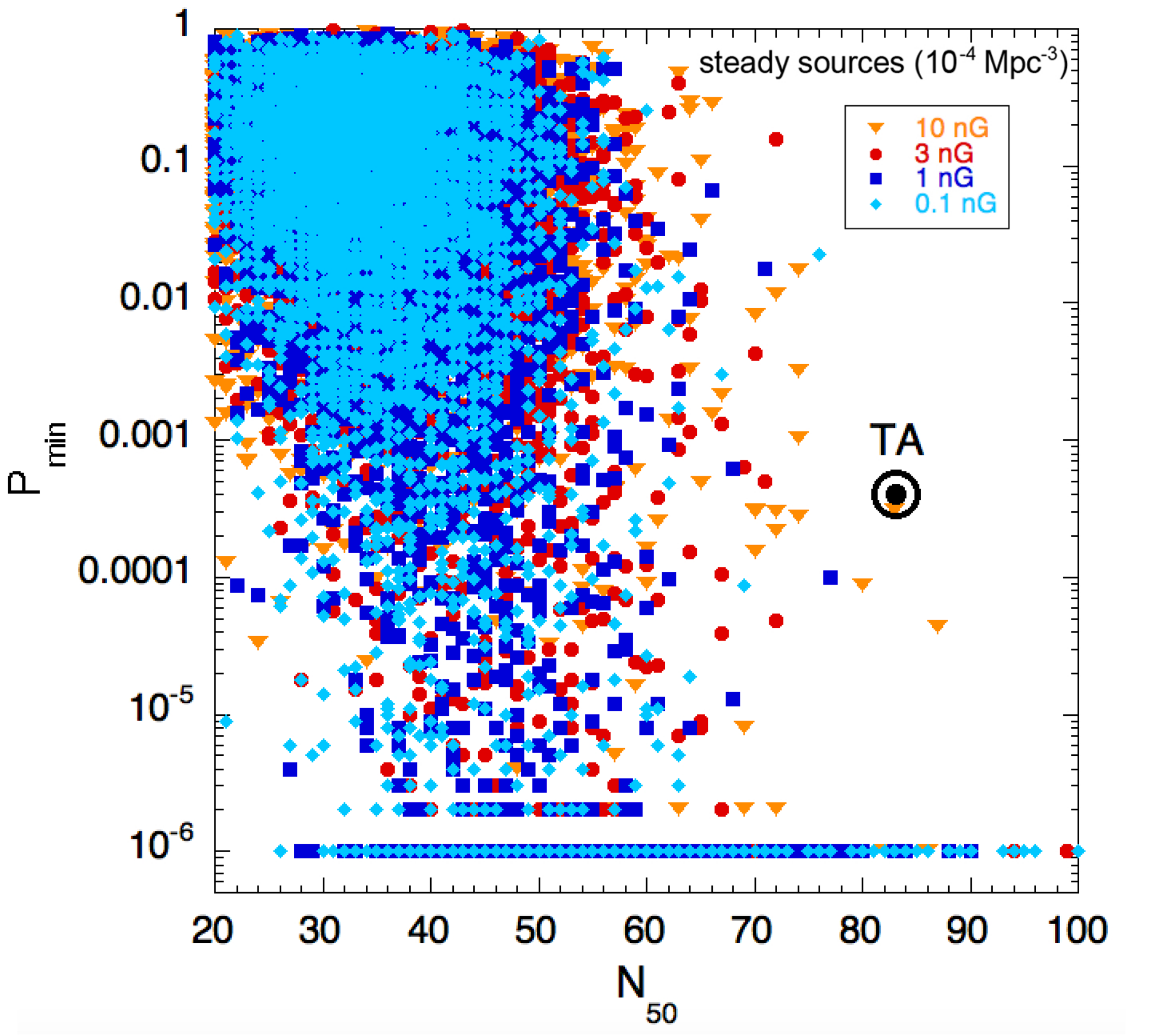}
\caption{Scatter plot of the values of $\mathcal{P}_{\min}$ vs. $N_{50}$ for all the TA-like data sets simulated in the steady source scenario, for a source density of $10^{-5}$ Mpc$^{-3}$ (left) and $10^{-4}$ Mpc$^{-3}$ (right), with 4 values of the magnetic field, as indicated. The position of the actual TA data set is indicated by the $\odot$ symbol. 
}
\label{fig:Correl2ptN50Steady40}
\end{center}
\end{figure*}

\subsubsection{A joined analysis of the spectrum and anisotropy constraints}\label{steadyres}
\begin{figure}[t!]
\begin{center}
\includegraphics[width=0.9\linewidth]{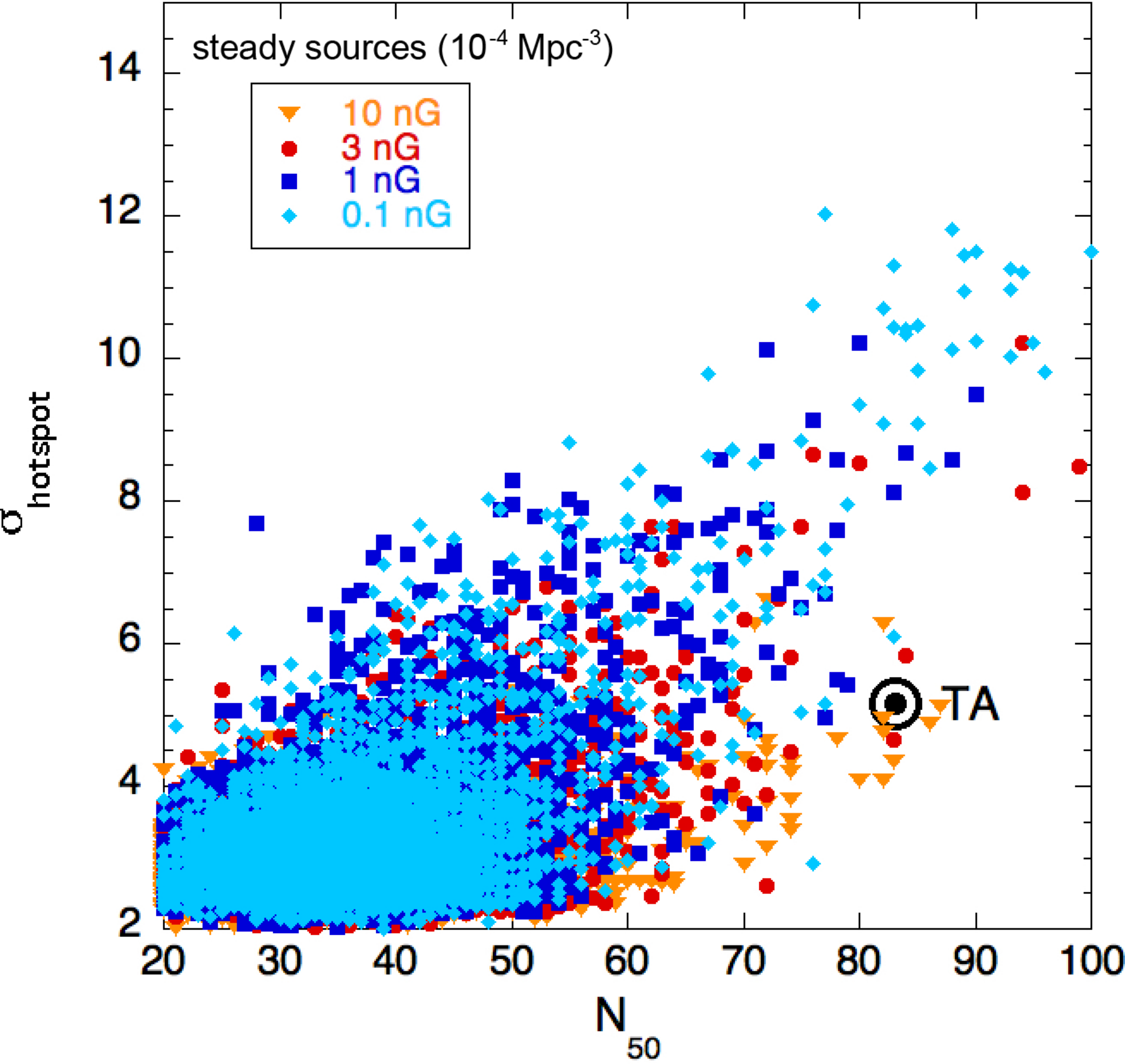}
\caption{Same as Fig.~\ref{fig:Correl2ptN50Steady40}, with the values of $\sigma_{\mathrm{hotspot}}$ vs. $N_{50}$, for a steady source density of $10^{-4}$ Mpc$^{-3}$.}
\label{fig:CorrelTopHatN50steady40}
\end{center}
\end{figure}

{\color{black} 
The different elements allowing the combined analyses of the spectrum and anisotropy constraints for the steady case are 
displayed in Figs. ~\ref{fig:Correl2ptN50Steady40} and \ref{fig:CorrelTopHatN50steady40}. 
Once more those scatter plots take into account only  realisations which have passed both the Auger 2-point and flux criteria (see Sect. \ref{TAanifluxtr}).}

{\color{black}
Fig.~\ref{fig:Correl2ptN50Steady40} depicts the scatter in the $N_{50}$ and $\mathcal{P}_{\min}$ plane. 
Similar comments as in the case of transient sources can be made. First, it is clear that TA data are very far from most simulated data sets. These have much lower values of $N_{50}$ than that measured by TA,  
and their anisotropy is smaller. 
The vast majority of the data sets that have a $\mathcal{P}_{\min}$ compatible with that of TA does not have a particularly strong excess in the high-energy flux. 
On the other hand, there is a small fraction 
of the data sets which do have a large flux excess, but almost all of them exhibit a much larger anisotropy than the TA data, with values of $\mathcal{P}_{\min}$ lower than $10^{-6}$, all the more for the lowest values of the EGMF. {\color{black} The left panel of Fig.~\ref{fig:Correl2ptN50Steady40} showing the $10^{-5}\,\rm{Mpc}^{-3}$ density case appears indeed very similar to the transient case shown in Fig.~\ref{fig:Correl2ptN50Transient}. The situation is slightly better for the $10^{-4}\,\rm{Mpc}^{-3}$, which shows that the few data sets yielding a strong flux excess (say, $E_{83}\geq70$ EeV) in the largest EGMF case do not necessarily show prohibitive values of $\mathcal{P}_{\min}$ and lie in the vicinity of the TA data point. }

}

{\color{black}
Fig.~\ref{fig:CorrelTopHatN50steady40} that depicts the scatter in the 
 $\sigma_{\mathrm{hotspot}}$ and $N_{50}$ plane,
shows that most of the data sets compatible with the TA flux excess above 50~EeV are much more strongly clustered than the TA data, with too large values of $\sigma_{\mathrm{hotspot}}$ (upper right part of the scatter plot). However, it tends to extend a bit more towards the data point, mostly in the case of high magnetic fields. Unlike, the case of transient sources, a sizable fraction of these data sets lying in the vicinity of the data point do not show a much more significant clustering at lower angular scales.   
%


}


\subsection{Quantitative discussion}
\label{sec:table}
{\color{black}A can be seen from the the scatter plots, the TA point is consistently far away from the bulk of the simulated data sets, for all values of the EGMF and source density. However a few realisations in the steady case 
give rise to a combination of a flux excess and anisotropy properties which are relatively close to the Auger and TA data. In this subsection, we explore various regions in the vicinity of the TA point in the 3D ($N_{\rm{50}},\, \mathcal{P}_{\min},\, \sigma_{\mathrm{hotspot}}$) space to quantify the above qualitative discussions (see Table~\ref{tab:summaryTab}). We consider only steady  sources with a  density of $10^{-4}\,\rm{Mpc}^{-3}$, as this case turns out to be the most favorable.

If one only asks whether the explored models can produce data sets which are at least as anisotropic\footnote{The results are almost independent of the question whether the anisotropy is measured using 
$\mathcal{P}_{\min}\leq 4\,10^{-4}$ or $\sigma_{\mathrm{hotspot}}\geq 5.1$.} and have at least as large a flux excess as the TA data set (while still being compatible with the Auger data in the Auger part of the sky), one finds that
about $\sim0.6$\% of the data sets satisfy the requirement for the lowest EGMF value ($B_{\mathrm{EGMF}}=0.1$~nG).  
The fraction is even lower for larger magnetic fields (see  Table~\ref{tab:summaryTab} lines \#1 to \#2). 
Results with $\sigma_{\mathrm{hotspot}}\geq 5.1$ instead of $\mathcal{P}_{\min}\leq 4\,10^{-4}$ are almost similar. However, as stressed in Sect. \ref{TAanifluxtr}, the clustering criterion is not totally redundant, as shown in line \#3 of the table, where the condition  $\sigma_{\mathrm{hotspot}}\geq 5.1$ has been added.

\begin{table*}[ht!]
\begin{center}
\begin{tabular}{|c|c|c|c|c|c|c|c|c|}
\hline
&combination of criteria & 10 nG & 3 nG & 1 nG & 0.1 nG \\
\hline
\hline
\#1 & $N_{50}\geq83$ & $0.050$ & $0.17$ & $0.067$ & $0.57$\\
\#2 & $N_{50}\geq83$, \,$\mathcal{P}_{\min}\leq 4\,10^{-4}$& $0.050$ & $0.17$ & $0.067$ & $0.57$ \\
\hline
\hline
\#3 &$N_{50}\geq83$, \,$\mathcal{P}_{\min}\leq 4\,10^{-4}$, \,$\sigma_{\mathrm{hotspot}}\geq 5.1$& $0.017$ & $0.15$ & $0.067$ & $0.57$ \\
\#4 &$N_{50}\geq83$, \,$10^{-6}<\mathcal{P}_{\min}\leq 4\,10^{-4}$, \,  $5.1\leq \sigma_{\mathrm{hotspot}}\leq8$& $0.017$ & $0$ & $0$ & $0$ \\
\#5 &$N_{50}\geq83$,  \,$4\,10^{-5}\leq \mathcal{P}_{\min}\leq 4\,10^{-3}$, \,  $4.6\leq \sigma_{\mathrm{hotspot}}\leq5.5$& $0.017$ & 0 & 0 & 0  \\
\#6 &$N_{50}\geq83$,  \,$4\,10^{-5}\leq \mathcal{P}_{\min}\leq 4\,10^{-3}$, \,  $4.1\leq \sigma_{\mathrm{hotspot}}\leq5.9$& $0.033$ & 0 & 0 & 0  \\
\#7 &$N_{50}\geq83$,  \,$4\,10^{-6}\leq \mathcal{P}_{\min}\leq 4\,10^{-2}$, \,  $4.1\leq \sigma_{\mathrm{hotspot}}\leq5.9$& $0.033$ & 0 & 0 & 0 \\
\#8 &$N_{50}\geq74$,  \,$4\,10^{-6}\leq \mathcal{P}_{\min}\leq 4\,10^{-2}$, \,  $4.1\leq \sigma_{\mathrm{hotspot}}\leq5.9$& $0.050$ & $0$ & 0.017 & 0  \\
\#9 &$N_{50}\geq70$,  \,$4\,10^{-6}\leq \mathcal{P}_{\min}\leq 4\,10^{-2}$, \,  $4.1\leq \sigma_{\mathrm{hotspot}}\leq5.9$& $0.083$ & $0.017$ & 0.017 & 0  \\
\#10 &$N_{50}\geq65$,  \,$4\,10^{-6}\leq \mathcal{P}_{\min}\leq 4\,10^{-2}$, \,  $4.1\leq \sigma_{\mathrm{hotspot}}\leq5.9$& $0.13$ & $0.033$ & $0.033$ & 0 \\
\#11 &$N_{50}\geq60$,  \,$4\,10^{-6}\leq \mathcal{P}_{\min}\leq 4\,10^{-2}$, \,  $4.1\leq \sigma_{\mathrm{hotspot}}\leq5.9$& $0.13$ & $0.07$ & $0.07$ & 0.05  \\
\#12 &$N_{50}\geq55$,  \,$4\,10^{-6}\leq \mathcal{P}_{\min}\leq 4\,10^{-2}$, \,  $4.1\leq \sigma_{\mathrm{hotspot}}\leq5.9$& $0.18$ & $0.23$ & $0.20$ & $0.17$  \\
\#13 &$N_{50}\geq40$,  \,$4\,10^{-6}\leq \mathcal{P}_{\min}\leq 4\,10^{-2}$, \,  $4.1\leq \sigma_{\mathrm{hotspot}}\leq5.9$& $0.35$ & $1.3$ & $1.1$ & $1.1$ \\
\hline
\end{tabular}
\\
 \end{center}
\caption{Percentage of simulated data sets satisfying  various combinations of criteria of compatibility with the TA data, for the steady source scenario assuming a density of $10^{-4}$ Mpc$^{-3}$. Each column corresponds to a different assumption on the EGMF, as indicated. The title of the lines refer to the different combinations of criteria (see text). Only the data sets which also satisfy the Auger criteria are indicated here. In the case of the different $\mathcal{P}_{\min}$ ranges (lines~\#5 to 13), only data sets for which $\theta_{min}\geq 15^\circ$ are considered.}
\label{tab:summaryTab}
\end{table*}%

On one hand, the scatter plots (see Figs. \ref{fig:Correl2ptN50Steady40} and \ref{fig:CorrelTopHatN50steady40}) show that in most cases, these data sets 
have values of $\mathcal{P}_{\min}$ lower than $10^{-6}$ and/or  values of $\sigma_{\mathrm{hotspot}}$ larger than 8. This means that those data sets 
show very significant anisotropies at small angular scales, contrary to what is observed. 
As a matter of fact, if among these data sets, one selects only those with $\mathcal{P}_{\min} > 10^{-6}$ and $\sigma_{\mathrm{hotspot}} \le 8$, only one data set (out of  6000) survives, at the largest EGMF value (10 nG), as shown in line~\#4 of the table.

On the other hand, a few data sets with anisotropies slightly less significant than indicated by the TA data are found in the scatter plots, and those may 
represent a reasonable description of the observations. We thus explored various regions around the TA data point to determine which sets of model parameters are most likely to produce data sets relatively close to the data. 
The results are shown in lines \#5 to 13 of Table~\ref{tab:summaryTab}, which shows the percentage of data sets entering boxes of increasing volume around the TA data point. In line~\#5, we require a flux excess at least as strong as that of TA, as well as pre-trial probabilities of both anisotropy observables within a factor of 10 of that found in the TA data (which translates as values of $\sigma_\mathrm{hotspot}$ between 4.6 and 5.5, { and values of $\mathcal{P}_{\min}$ between $4\,10^{-5}$ and $4\,10^{-3}$}). In line \#6, this range of pre-trial probabilities is further extended to a factor of 100 for $\sigma_{\mathrm{hotspot}}$. In line \#7, the range for $\mathcal{P}_{\min}$ is also extended to 100 times larger or smaller than the TA value. In lines \#8 to \#13, we keep the latter ranges for the anisotropy observables, and allow for smaller and smaller flux excesses from $N_{50} \geq 74$ (1$\sigma$ away from the observed value of 83) down to  $N_{50} \geq 40$.

As can be seen, only a few data sets, all corresponding to $B_{\mathrm{EGMF}}=10$~nG, enter the most restrictive boxes:  1 data set (over 6000) in the box corresponding to line \#5, and 2 in the boxes of lines \#6 and \#7). Loosening the requirement on the flux excess, with lower values of the $N_{50}$ lower bound (lines \#8 and \#9) allows the first few data sets for lower EGMF values (1 and 3~nG) to appear in the boxes. However, the bulk of the data sets passing the cuts is still found for the largest EGMF value.  Data sets simulated with a 0.1~nG EGMF are found to enter an enlarged box around the TA data point only at line~\#11, i.e. with a lower bound on $N_{50}$ set at 60 events (which is $\sim 2.5\sigma$ away from the observed value). The number of data sets included in the box only becomes roughly independent of the assumed EGMF at line~\#12, i.e. for $N_{50} \geq 55$ only: it then concerns $\sim 0.2\%$ of the data sets. Finally, line~\#13 shows that, if one does not particularly request a flux excess, i.e. we accept all data sets with $N_{50}$ larger than the median of its distribution (i.e. $\sim 40$), the extended anisotropy criteria are passed by about 1\% of the data sets, except for the largest value of the EGMF, for which the probability is about $\frac{1}{3}$ lower. 

Based on a similar quantitative discussion, the lower source density case ($10^{-5}\,\rm{Mpc}^{-3}$) appears even less favorable. Indeed, data sets calculated  with the largest EGMF suffer from a stronger magnetic horizon effect and the data sets showing a strong flux excess are mostly to be discarded due to the "Auger flux" criterion. As a result no data set (including all the considered EGMF values) are able to populate the most restrictive boxes. The most compatible data sets are only found in the enlarged box corresponding to line~\#9, in a lower percentage ($\sim 0.03\%$) than in the $10^{-4}\,\rm{Mpc}^{-3}$ case.

Likewise, in the transient case, due to the very low probability to produce a strong flux excess for large values of the EGMF and the very significant anisotropies systematically associated with strong flux excesses in the lowest EGMF cases, data sets are found at best to enter enlarged boxes corresponding to line~\#11 in Table.~\ref{tab:summaryTab}.

}

\section{Summary and discussion}
\label{sec:discussion}

In this paper, we have addressed the compatibility between the Auger and Telescope Array data on UHECRs. We first showed that, taken at face value, the two energy spectra don't appear to be mutually compatible. The TA spectrum cannot be considered a mere statistical fluctuation of the Auger spectrum, and vice versa, even if one allows for a global shift in the relative energy scale of the two experiments. In particular, the integrated flux measured by TA above 50~EeV is significantly larger than that of Auger. This cannot be accounted for if the underlying UHECR flux is approximately the same in all direction, as the current anisotropy analyses reported by Auger suggests. However, the TA collaboration has reported a possible hotspot with an angular scale around 20$^{\circ}$ in the Northern hemisphere, in a part of the sky that is not observed by Auger. One may thus ask  whether both features -- a cluster of events and an excess in the flux at the highest energies -- could be two complementary manifestations of a single reality: the presence of a very bright source in the Northern sky.

Assuming that the current Auger and TA data are indeed representative of the actual characteristics of the UHECRs in their respective parts of the sky, we investigated the possibility, for a given source model, to satisfy the various observational constraints. We considered a wide range of astrophysical scenarios, including transient sources or steading sources, as well as different source densities and different values of the EGMF.

Simple analytic estimates show  that such a flux excess could typically occur at most in a few percent of the cases, either in the transient source scenario or in the steady source scenario. 
However, the flux excess needed to account for the difference between the Auger and TA spectra corresponds to a number of events which is much larger than the number of events in the TA hotspot. We estimated that if the flux excess is to be explained by the contribution of one dominant source, that source should contribute around $45\pm 6$ events in total, while the number of events in the so-called TA hotspot represents only $\sim 21\%$ of the total number of events including some possible background. Besides, it is striking that the highest energy events are not present in the hotspot region itself. In particular, in the initial report of the presence of an intermediate scale anisotropy in \cite{TAPaperOnHotspot}, none of the 12 events observed by TA above $10^{20}$~eV (7 if the energy is rescaled downward by 13\%) can be found within 20$^{\circ}$ of the hotspot center. 
This can be explained within our scenarios by simply noting that the rigidity of the highest energy particles is in fact smaller than those at intermediate energies, due to the change in composition. More specifically, CNO nuclei at 60~EeV have a rigidity twice as large as Fe nuclei at 100~EeV. The highest energy events might thus very well (and are actually expected to) be deflected more than the particles between 50 and 60~EeV, say. 

The analytical estimates also suggested that the angular size of the hotspot associated to the flux excess is too small compared to the one observed by TA. In this respect, it is interesting to note that transient sources suffer from a general problem: larger deflections, as needed to fit the angular extension of the dominant source, also imply larger spreads in the particles arrival time, which in turn reduce the apparent flux of the source, and thus makes it even less likely for a source to contribute a large fraction of the total UHECR flux. Steady sources, on the other hand, do not suffer from this problem, since their apparent flux does not depend on the time spread of the particles, but only on its distance. Larger magnetic fields, at least in the direction of the source, might thus in principle increase its apparent angular size, without reducing its flux.

Our detailed numerical simulations took into account the various effects influencing the propagation of the UHECRs, including energy losses, photodissociation in the case of nuclei, and deflections by the intervening magnetic fields, around the source, in the intergalactic medium and in the Galaxy. A remarkable feature that we noticed is that according to the representative model of the GMF by \cite{JF12}, UHECR propagation is  significantly different in the Northern and Southern skies. The Galactic magnetic deflections are smaller at the North and hence the angular spreads and time delays are typically smaller for particles in the TA sky than in the Auger sky.

{\color{black}
Our results are summarised in Figs. \ref{fig:Correl2ptN50Transient}, \ref{fig:PminNsigmaTransient}, \ref{fig:correlSigmaN50} for the transient source scenario and in Figs. \ref{fig:Correl2ptN50Steady40}, \ref{fig:CorrelTopHatN50steady40}  for the steady source scenario. 
Overall, we find that transient sources are essentially incapable of reproducing the data. } {\color{black}The steady source scenario does not appear prima facie to be very favorable either. 
However, steady sources are  more likely than  transients to account  for data sets with general features reminiscent of the Auger and TA data. This difference between the two models arise   because of the possibility to accept larger particle deflections, to attenuate the strong anisotropy usually produced by a strongly dominating source. Large values of the EGMF are needed in this case. 
A clear requirement of such solutions is the necessity to have a very nearby source, within $\sim 20$ Mpc as can be inferred from Figure \ref{fig:distances}), in order to obtain a large flux excess in some part of the sky.  Indeed nearby sources such as M82 have been suggested as the origin of the TA hotspot \citep{He16,Pfeffer16}. However, none of these nearby sources is within the Ursa Major hotspot itself. Models based on such a source  have to demonstrate that the large ($>20^o$)  average deflection required is indeed possible and that they do not produce too significant clustering at lower energies.  

}

Concerning the underlying assumptions, the scenarios studied here rely on some generic features of the GRB source model developed in \cite{Globus15a}. In particular, we used the same source composition and energy spectra, as well as the same distribution of relative luminosities at the sources. We argued and verified whenever possible that these specificities are not likely to modify significantly the conclusions of the study. Indeed, as far as the composition is concerned, it was shown previously to satisfy the current observational constraints derived from the Auger data. We note than an even heavier composition than the one we considered could in principle help attenuating the anisotropies, without reducing too much a possible flux excess in the TA sky, if the dominant source is far from the region of the sky observed by Auger, and most of its flux remains confined within the Northern equatorial hemisphere. 
However, assuming that the composition is almost at its heaviest  already at 50 EeV would make it more difficult to account for the the fact that the highest energy events are not within the TA's hotspot  while their number is also clearly in excess with respect to the Auger data. In the same line of reasoning, let us note that if the presence of a hotspot in TA dataset is confirmed with larger statistics, the study of the energy evolution of its significance will be critical in order to constrain the source composition and especially the presence of protons at lower energy. Indeed, \cite{Lemoine09}, pointed out that  a cluster 
of events due to cosmic-rays nuclei with charge $Z$ above an energy $E$, implies a  
clustering  of 
UHECRs with the same rigidity but lower $Z$ at the same spot. 
The  statistical significance of the lower energy enhancement depends on the details of the source composition and spectra.  We encourage the TA collaboration to search for such an enhancement and report on its existence or lack of. 


The last assumption is concerned with the intrinsic luminosity  distribution of  the sources. Clearly,  as long as the UHECR sources are not known, and the acceleration process is not identify, one cannot make any reliable statement about it. We assumed the luminosity distribution taken from the  GRB source model. However, scenarios based on standard candles or on the contrary with a much wider luminosity distribution are in principle possible. From the point of view of producing a strongly dominating source, our assumptions lead a larger variance in the observed spectra than the standard candle scenario for a given inferred source density. Now, if the luminosity distribution were actually larger, it would increase the cosmic variance, with some realisations having their closest source particularly bright, or on the contrary only weak sources among the most nearby ones. This, however, would generally not produce an effect very different from what could be expected in the case of a lower source density, where fewer sources contributing a larger flux individually. 


We note that the transient sources considered in this paper have been assumed to emit UHECRs isotropically, although a beamed emission is probable. 
As we pointed earlier the isotropic assumption adopted here is in fact optimistic in term of the probability to produce a significant flux excess since beamed sources would be more numerous, have smaller intrinsic luminosities and any angular spreading in the vicinity of the sources would reduce the effective flux emitted in the direction of the observer, and thus reduce the probability of a large flux excess. Another specificity of our treatment of transient sources is that we applied an additional time delay in the propagation of the particles as if the sources were located in an environment similar to the that of the Sun, assuming a host galaxy with magnetic properties like ours. In principle, the source might be located in a strongly magnetized environment. This would further reduce the possibility to observed a large flux excess in the case of transient sources, by implying yet additional time delays. If these time delays are long enough, it may even turn transient sources into steady sources. This possibility appears extremely unlikely in a galactic magnetic environment similar to ours (at least with the Galactic magnetic field model we used throughout our study) because the typical GRB rate per galaxy ($\sim 10^{-5}\,\rm yr^{-1}$, assuming a beaming factor $f \simeq 100$) is too low for the UHECR signal of different GRB to significantly overlap in time. Starburst galaxies which are known to harbor much stronger magnetic fields than our own Galaxy (see for instance \cite{Beck15} for a review) may appear as a better candidate for such an overlap between successive GRBs to take place. Let us note however that this possibility remains at present speculative, since there is no strong established connexion between starburst galaxies and GRB explosion in the local universe (see \cite{Japelj16, Vergani16} for recent accounts). 

Finally, from the point of view of the energy scale, we have chosen to rescale  TA energy scale downward by 13\% while we could have equivalently chosen to rescale Auger energy scale upward by the same amount. This choice was dictated both by the fact that Auger claimed systematic on the energy scale (14\%) is lower than that of TA (22\%) and also because at this energy scale the experimental value of $E_{231}=52$~EeV corresponds quite closely the median value found in our simulations for most of our  hypotheses on the extragalactic magnetic field, the rate or density of the sources. By choosing to rescale Auger energy scale by 13\% upward we would thus have discussed the excess of events in TA dataset above 57 EeV rather than above 50 EeV. While it intuitively obvious that the cosmic variance (and as a result the probability to observe a significant difference between TA and Auger number of events) is expected to increase with the energy, this effect should be quantitatively minor for a 13\% shift in energy, all the more in the case of a composition getting heavier in this energy range. As soon as it remains moderate a global shift of the energy scale should thus not affect significantly our conclusions and appear to be much less relevant than energy dependent systematic errors for our present discussion. Since it is precisely the conjunction between a large flux excess and a moderate anisotropy (the latter being reproduced in $\sim 1$\% of the simulated data sets) in the northern hemisphere which turned out to be particularly challenging to account for, the discovery of any energy dependent systematic effect allowing to reduce the high energy flux difference between the two hemispheres would alleviate the constraints imposed by current observations.

As conclusion, based on the results presented in this paper, the current features of the Auger and TA data, if confirmed by future observations, may either point towards rather unconventional astrophysical scenarios, or to a very particular situation in the universe, where we happen to be very close to an intense steady source, affected by particularly large magnetic deflections. Additional data, especially with reduced or ``homogeneous'' systematic uncertainties, allowing for a direct comparison between the Northern and Southern sky, with similar statistics, would definitely be of great help to address this issue. This could be provided by a space-based experiment like envisioned by the JEM-EUSO collaboration \citep{JEMEUSO}.

\section*{Acknowledgements}
We thank Reiner Beck for very useful comments about the Galactic magnetic field. NG thanks David Eichler for inspiring discussions about UHECRs for more than a year. DA, CL and EP are indebted to Benjamin Rouill\'e d'Orfeuil for important contributions to the initial version of the sky map drawing procedure and analysis. We also thank Hiroyuki Sagawa for providing important information on the Telescope Array data sets.
This research was supported by I-Core CHE-ISF center of excellence of research in astrophysics and by an Israel Space Agency grant (NG \& TP), by the Lady Davis foundation (NG), and by the UnivEarthS Labex program at Sorbonne Paris Cit\'e (ANR-10-LABX-0023 and ANR-11-IDEX-0005-02) (CL).

\appendix
\section*{Simulated data sets compatible with both TA and Auger}
\label{sec:exampleofrealisation}

\begin{figure}[t!]
\begin{center}
\includegraphics[width=0.6\linewidth]{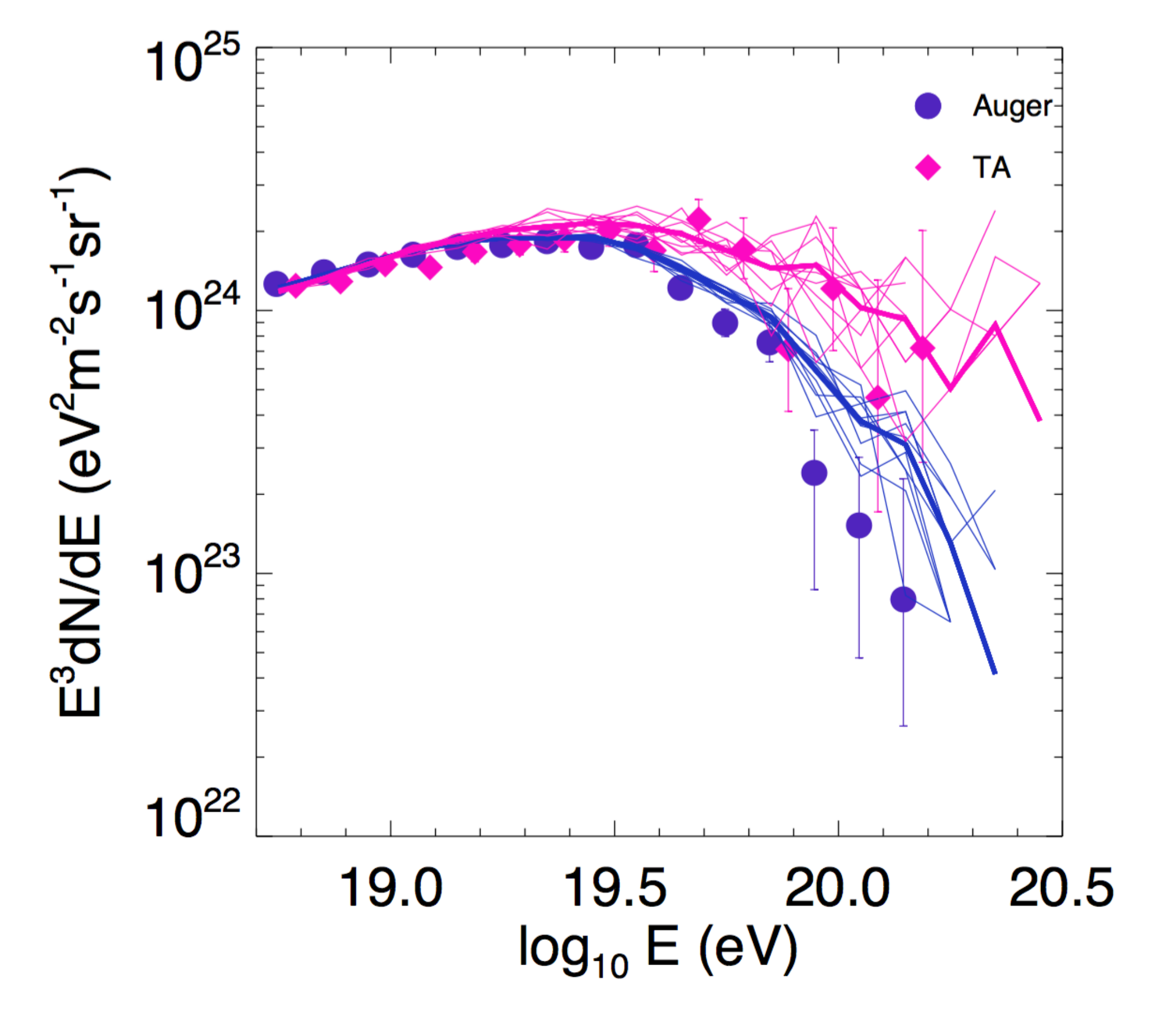}
\caption{Spectra of the 10 data sets (thin lines) built from a realisation of the steady source scenario (with a source $10^{-4}$ Mpc$^{-3}$ and an EGMF of 10~nG) which jointly satisfies the spectrum and anisotropy criteria of compatibility with the observations, together with the Auger and TA data points. The thick lines correspond to the average spectra of the 10 data sets.} 
\label{fig:spectreOKSteady}
\end{center}
\end{figure}

\begin{figure*}[t!]
\begin{center}
\includegraphics[width=0.5\columnwidth]{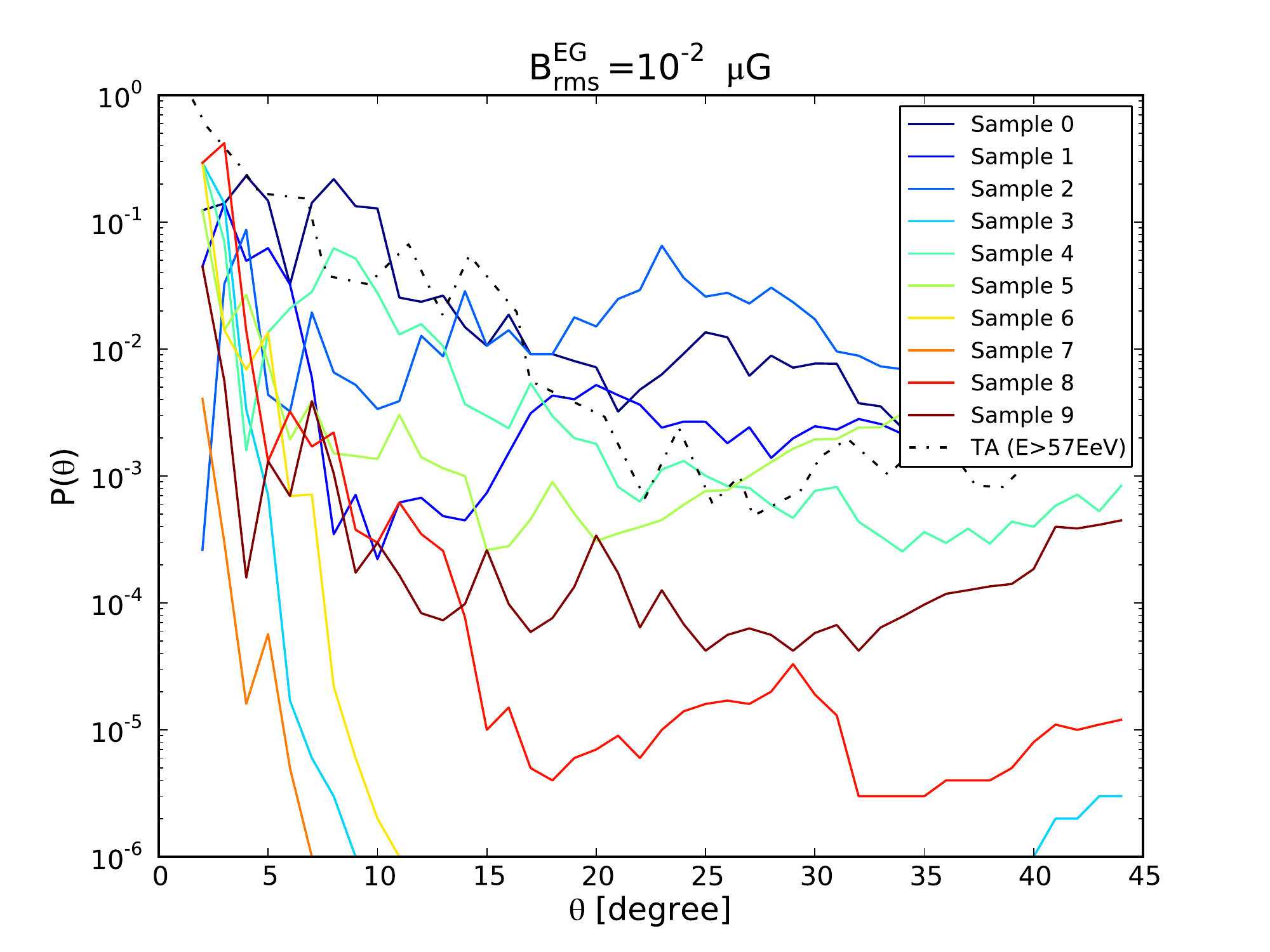}\includegraphics[width=0.5\columnwidth]{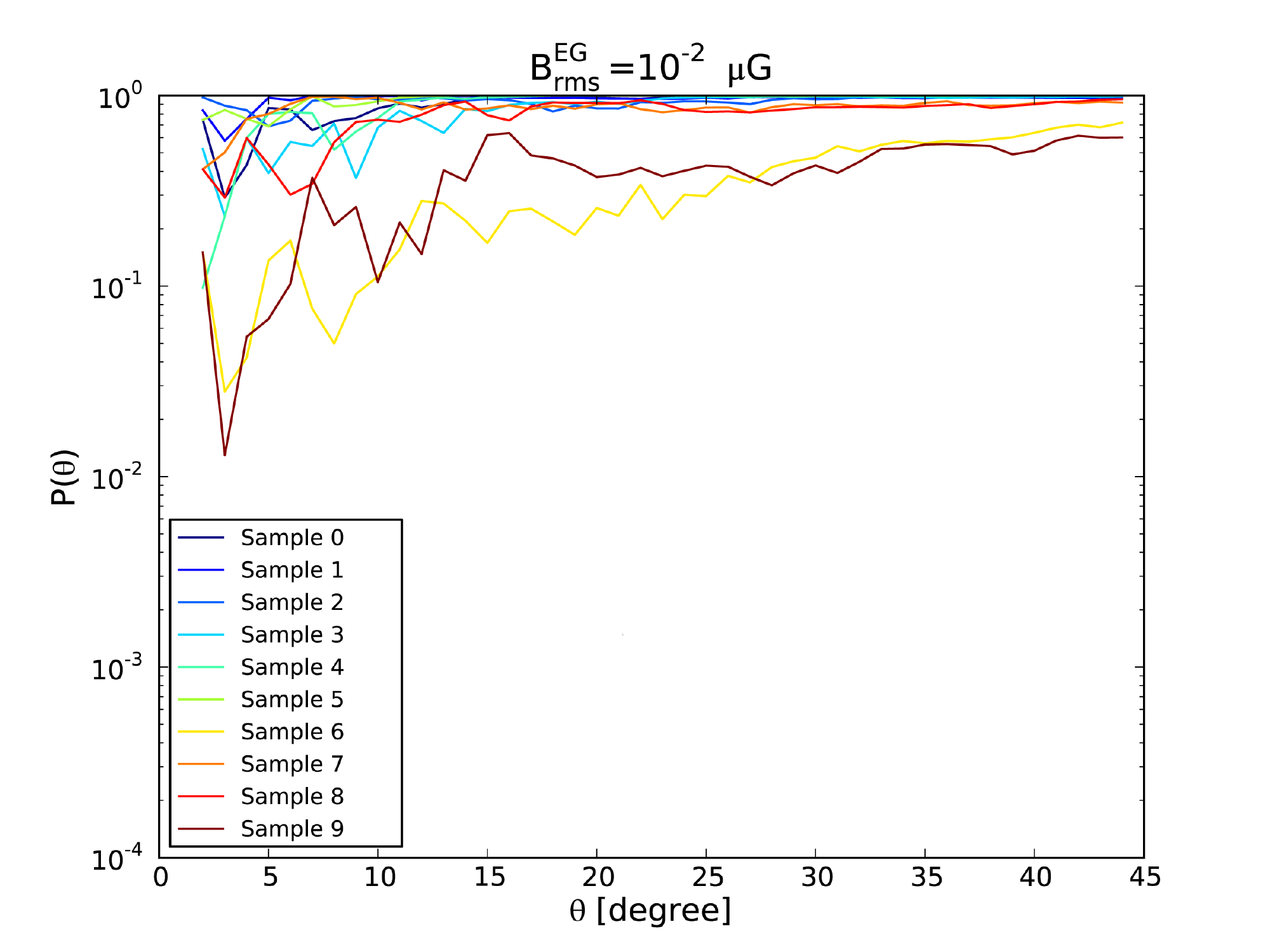}
\caption{Two-point correlation functions of the 10 data sets obtained with the only realisation of the steady source scenarios which jointly satisfies the spectrum and anisotropy criteria of compatibility with the actual data (see text). Left: TA-like simulated data sets. The dashed line shows the 2-point correlation function of the actual TA data. Right: Auger-like simulated data sets.}
\label{fig:2ptOKSteady}
\end{center}
\end{figure*}

\begin{figure*}[t!]
\begin{center}
\includegraphics[width=0.5\columnwidth]{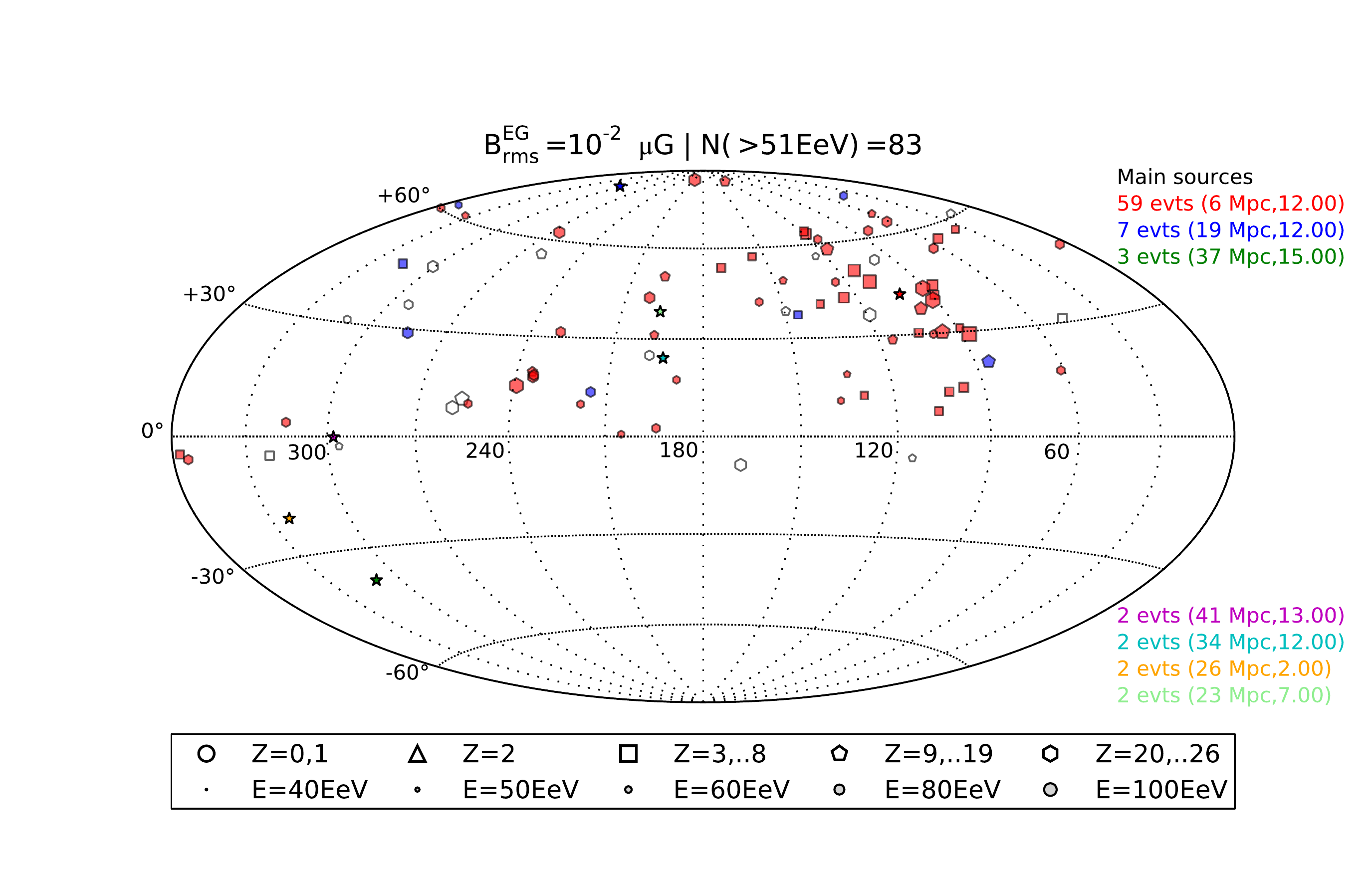}\includegraphics[width=0.5\columnwidth]{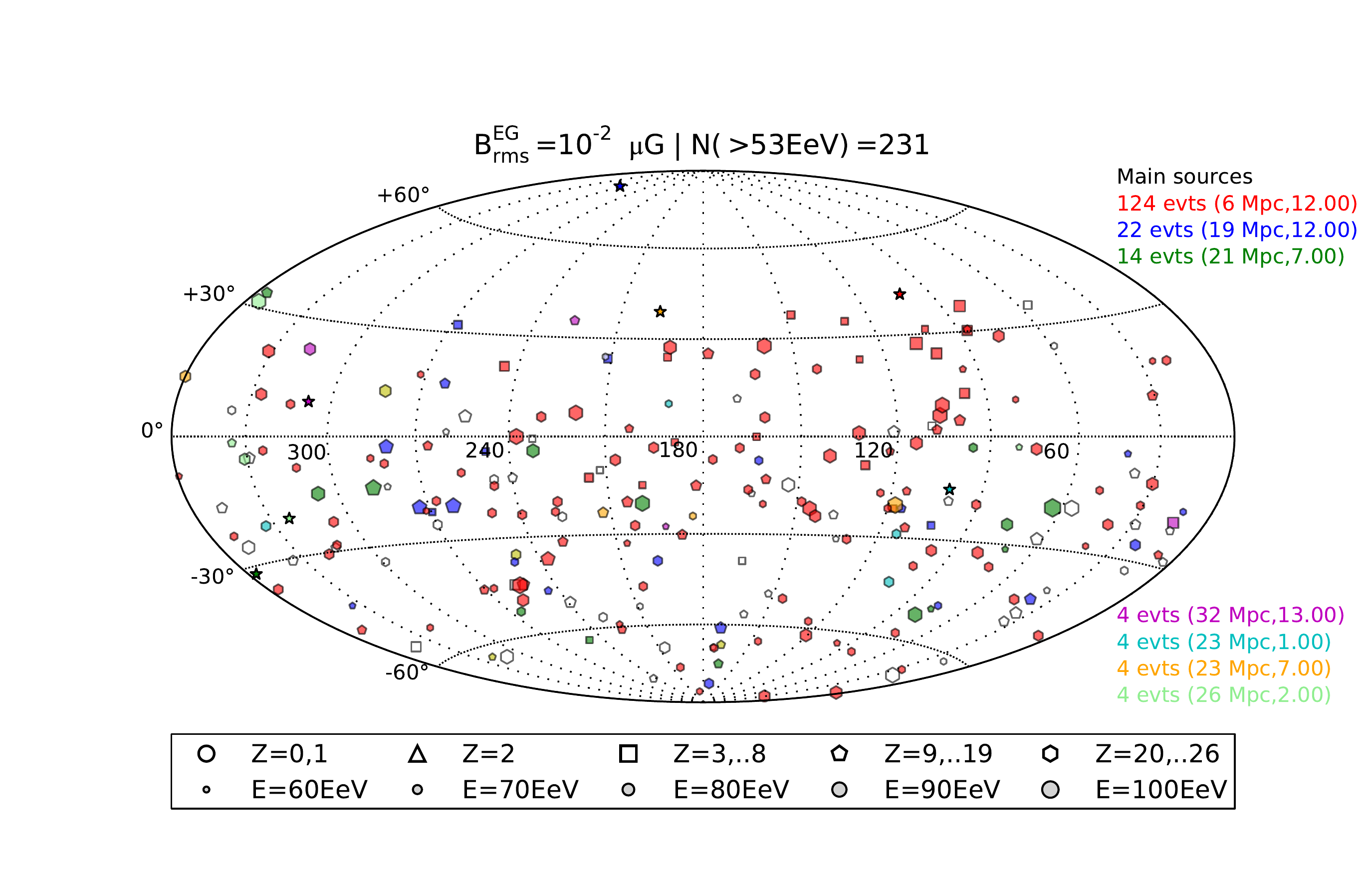}
\caption{TA-like (left) and Auger-like (right) sky maps, in equatorial coordinates, corresponding to sample~9 of the steady source realisation shown in Figs.~\ref{fig:spectreOKSteady} and~\ref{fig:2ptOKSteady}. The different symbols correspond to the different groups of species contributing to the flux (see the legend) and the size of the symbols being proportional to the logarithm of the energy of the individual events). The different colors allows to distinguish the contribution of the brightest sources (their distance and contribution to the flux is indicated on the right side of the map with the corresponding color).}
\label{fig:mapsOKSteady}
\end{center}
\end{figure*}

{\color{black}
In the following, we describe the characteristics of one of our realisation that can account satisfactorily to Auger and TA. It includes 3 data sets enclosed respectively in boxes~\#5 (satisfying also the criteria of line~\#4), 9 and 10 defined in Table.~ \ref{tab:summaryTab}.

As can be seen on Fig.~\ref{fig:spectreOKSteady}, this  realisation (obtained for the steady source scenario assuming a $10^{-4}\,\rm{Mpc}^{-3}$ source density and a 10 nG EGMF)  has distinctly different spectra in the Auger sky and in the TA sky, both very similar to those of the actual data. Essentially each of the 10 data sets simulated from this realisation have a suitable spectrum, with values of $N_{50}$ which are all between 70 and 87, except one, which has $N_{50} = 67$. The average value is 74.

Regarding the criterion on $\mathcal{P}_{\min}$, we show in Fig.~\ref{fig:2ptOKSteady} the 2-point correlation functions of the 10 data sets of the realisation. All the Auger-like data sets (right panel) are seen to be compatible with isotropy, in agreement with the Auger data. As for the TA-like data sets (left panel), in 4 cases out of 10, the 2-point correlation function is similar to that of TA (represented by the dashed line). Another one is also similar, but with lower probabilities at small angles. A sixth one may be considered marginally compatible, while the last 4 are clearly too  anisotropic.

Finally, Fig.~\ref{fig:mapsOKSteady} shows the sky maps corresponding to the data set entering in all the boxes shown in Table.~ \ref{tab:summaryTab} including the most constraining one, namely sample 9. It has the following values of the parameters: $N_{50} = 87$, $\sigma_{\mathrm{hotspot}} = 5.1$, and $\mathcal{P}_{\min} = 4\cdot10^{-5}$. The color code in the figure allows us to identify the events coming from the same source. As can be seen, in the TA-like sky map (top panel) 59 events are coming from one single source, located at a distance of 6~Mpc, with a luminosity in cosmic rays above  $10^{18}$ eV of $\sim 3\cdot10^{41}$ erg.s$^{-1}$.
The anisotropy of this data set remains however moderate, mostly thanks to the large value of the EGMF, with events spreading over the whole sky. The second most intense source only contributes 7 of the 83 highest energy events, also distributed all over the map. Although the Auger-like sky map (lower panel) looks indeed compatible with isotropy (as was indicated by the 2-point correlation function), it is interesting to note that the dominant source in the TA sky is also vastly dominant in the Auger sky, contributing no less than 124 of the 231 highest energy events. However, these events are indeed spread throughout the whole sky, so the anisotropy remains low. It should also be noted that the second most intense source in this Auger-like data set only contributes 22 events. This explains why the largely dominant source, visible in both hemispheres, is not leading to a UHECR flux that is too large compared with the high-energy flux measured by Auger. In other words, this particular realisation of the model actually corresponds to an otherwise downward fluctuation of the flux in the Auger-like sky, compensated by the large contribution of the most intense source in the sky to the overall flux in all directions.}

\end{document}